\documentclass[%
reprint,
amsmath,amssymb, aps,
prb,
]{revtex4-2}

\usepackage{graphicx}
\usepackage{dcolumn}
\usepackage{bm}
\usepackage{placeins}
\usepackage{float}
\usepackage{svg}
\usepackage{afterpage}
\usepackage{multirow}
\usepackage{textcomp}
\usepackage{lineno}
\usepackage{booktabs,array,calc}
\usepackage[utf8]{inputenc}
\usepackage[T1]{fontenc}
\usepackage{microtype}
\usepackage{xcolor}

\DeclareUnicodeCharacter{03A6}{\ensuremath{\Phi}}
\DeclareUnicodeCharacter{03C6}{\ensuremath{\phi}}
\DeclareUnicodeCharacter{03C0}{\ensuremath{\pi}}
\DeclareUnicodeCharacter{03BB}{\ensuremath{\lambda}}
\DeclareUnicodeCharacter{03B1}{\ensuremath{\alpha}}
\DeclareUnicodeCharacter{03B2}{\ensuremath{\beta}}
\DeclareUnicodeCharacter{03BC}{\ensuremath{\mu}}
\DeclareUnicodeCharacter{03C9}{\ensuremath{\omega}}
\DeclareUnicodeCharacter{03C7}{\ensuremath{\chi}}
\DeclareUnicodeCharacter{03A9}{\ensuremath{\Omega}}
\DeclareUnicodeCharacter{2260}{\ensuremath{\neq}}
\DeclareUnicodeCharacter{2265}{\ensuremath{\ge}}
\DeclareUnicodeCharacter{2264}{\ensuremath{\le}}
\DeclareUnicodeCharacter{226A}{\ensuremath{\ll}}
\DeclareUnicodeCharacter{00D7}{\ensuremath{\times}}
\DeclareUnicodeCharacter{2212}{-}
\DeclareUnicodeCharacter{2013}{--}
\DeclareUnicodeCharacter{2014}{---}
\DeclareUnicodeCharacter{2019}{'}
\DeclareUnicodeCharacter{2018}{`}
\DeclareUnicodeCharacter{201C}{``}
\DeclareUnicodeCharacter{201D}{''}
\DeclareUnicodeCharacter{00A0}{~}
\DeclareUnicodeCharacter{2009}{\,}
\DeclareUnicodeCharacter{FB01}{fi}
\DeclareUnicodeCharacter{FB02}{fl}

\begin{document}
\preprint{APS/123-QED}

\title{Superconducting parametric amplifiers based on arrays of unequal Josephson junction elements}
\author{Boris Chesca}\email{bchesca@oqc.tech}
\author{Tom Dixon}
\author{Narendra Acharya}
\author{Renuka D. Pothuraju}
\author{Jacob W. Dunstan}
\author{Connor D. Shelly}\email{cshelly@oqc.tech}
\author{Oscar W. Kennedy}\email{okennedy@oqc.tech}
\affiliation{Oxford Quantum Circuits, Thames Valley Science Park, Shinfield, Reading, UK}

\begin{abstract}
Impedance matching is a crucial feature of an amplifier used to increase its instantaneous bandwidth and is typically achieved using static LC resonators. Here we show an approach to impedance matching in parametric amplifiers using arrays of unequal Josephson-junction-based elements (rf/dc SQUIDs, SNAILs, etc). The elements act either as small-footprint flux-tunable inductors forming an impedance-matching circuit, or `active' amplifier elements which result in amplifier gain. We develop an analytical theory describing these systems, predicting a quasi-non-periodic amplifier frequency and gain modulation with applied flux and the broadening of gain bandwidths. We demonstrate this in devices using arrays of flux-focused SNAILs where flux biasing varies across the array. We find instantaneous bandwidths of 200~MHz with $>$17~dB of gain centred at $\sim$5.4~GHz in a device without a static impedance matcher. This broadening of amplifier bandwidth is stable with $>$100~MHz bandwidth at 17~dB gain across $>$1.2~GHz tunable bandwidth. Our analytical model captures flux dependence and qualitatively reproduces gain-bandwidths giving confidence in our understanding of these systems and allowing operational figures of merit to be optimised in future devices.

\end{abstract}
\maketitle

\section{Introduction}
Josephson Parametric Amplifiers (JPAs) are a common choice as first-stage, near-quantum-limited microwave amplifiers in superconducting-qubit readout
\cite{yamamoto2008flux,krantz2016single,white2023readout,joshi2025lumped,vijay2011observation},
quantum sensing \cite{hatridge2011dispersive}, microwave quantum optics \cite{roy2016introduction,roy2018quantum,aumentado2020superconducting}, and particle physics \cite{backes2021quantum,uchaikin2024josephson}. They are relatively straightforward to fabricate~\cite{ranzani2024broadband}, can be operated with low or near-quantum-limited added noise~\cite{yurke1988observation,yurke1989observation,roy2016introduction,roy2018quantum,aumentado2020superconducting,uchaikin2024josephson,white2023readout,kaufman2025simple} and using static impedance-matching circuits can realise moderate instantaneous bandwidths~\cite{mutus2014strong,ezenkova2022broadband,kaufman2023josephson,white2023readout}. JPAs are complemented by Josephson Travelling Wave Parametric Amplifiers (JTWPAs) \cite{aumentado2020superconducting} which find use in similar applications. JTWPAs offer superior GHz-scale instantaneous bandwidth~ \cite{obrien2014resonant,macklin2015near,white2015traveling,planat2020photonic} at the expense of more complicated fabrication, often including hundreds to thousands of individual Josephson junctions~\cite{macklin2015near,white2015traveling,zorin2017traveling,o2023simulating}, and can suffer from distributed insertion loss~\cite{zhao2019loss,chang2025josephson}.

Impedance matching circuits~\cite{naaman2022synthesis} made from distributed elements can often result in large circuit elements at the frequencies common for JPAs~\cite{mutus2014strong,ezenkova2022broadband,kaufman2023josephson,white2023readout}. Lumped element impedance matching circuits~\cite{joshi2025lumped} can be miniaturised by introducing new circuit elements; interdigitated~\cite{joshi2025lumped} or parallel plate capacitors and lumped element inductors~\cite{moskaleva2024lumped}. Here we introduce a new approach to impedance matching which uses arrays of unequal Josephson elements, the same circuit element that introduces non-linearity to the amplifiers. Some Josephson elements are tuned to amplifying working points, while others act as tunable inductors in a small footprint. 
This approach to impedance matching has the potential for miniaturised JPAs without introducing a dedicated impedance matching circuit and may either replace, or complement the effects of, static circuit-based impedance matching circuits.

The article is organized as follows: in section II we present a comprehensive analytical theory of JPAs incorporating arrays of $M$ unequal Josephson elements considering the general situation and the specific case of Superconducting Nonlinear Asymmetric Inductive eLements (SNAILs)~\cite{frattini2017wave,frattini2018optimizing,sivak2019kerr}. In section III, we present a representative set of measurements to demonstrate an improved performance of JPAs based on $M=3$ and $M=5$ unequal SNAIL arrays where the SNAILs are distinguished by varying their effective areas across the array. We implement this by fabricating additional flux-focusers directly integrated on some of the SNAILs in the array. We show that the analytical theory developed in section II qualitatively matches the experimental data and that this engineering approach realises 200~MHz bandwidth $>$17~dB gain centred at $\sim$5.4~GHz and that the increase in bandwidth is stable with over 100~MHz, 17~dB gain across a tunable bandwidth of $>$1~GHz.

\section{Arrays of unequal Josephson junction elements: theory}

Here we introduce an analytical model of parametric amplifiers based on arrays of $M$ unequal Josephson Junction Elements (JJEs). JJEs are circuit elements containing $N$ Josephson junctions, typically embedded into a superconducting loop. $N=1$ JJEs are known as rf (radio frequency) Superconducting Quantum Interference Devices (SQUIDs), $N=2$ are known as dc (direct current) SQUIDs and $N\ge 3$ elements include SNAILs. We assume that the system consists of $P$ (with $P\le M$) distinct JJEs and is therefore described by a $P$-dimensional potential. In practical devices, some JJEs, but not necessarily all, may differ from one another, meaning the superconducting phase drop splits unevenly across them. In order to derive an analytical description of these devices with arrays of unequal JJEs we start by considering the $P$-dimensional potential that describes the array (for the most inclusive potential see Appendix A). From this potential we can extract the Josephson inductance of the individual JJEs (partial derivatives), the device resonant frequency (inductance used in a circuit model) and the $g$ coefficients of the JPA Hamiltonian relevant for gain (Taylor expanding the potential).

\begin{figure}
\centering
\includegraphics[width=0.5\textwidth]{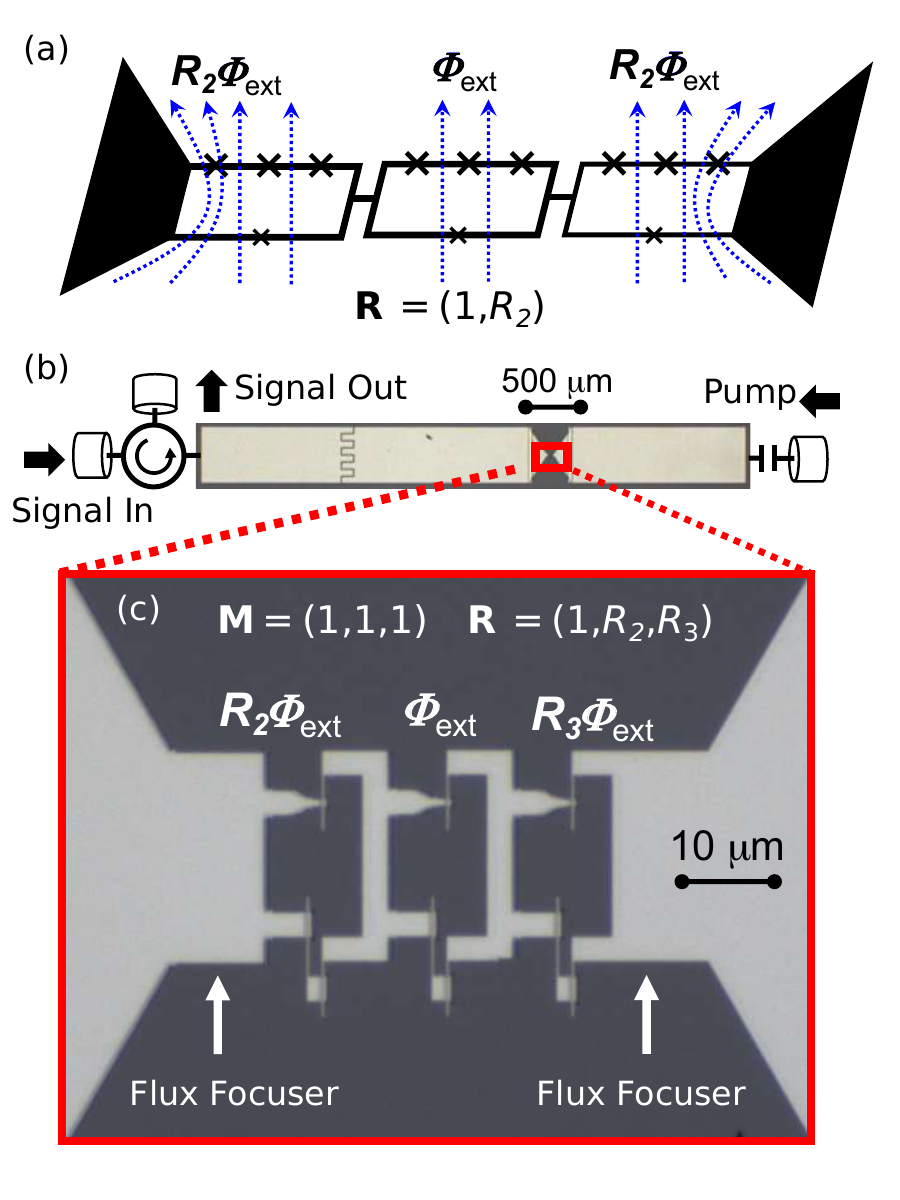}
\caption{(a) Schematics of an $\mathbf{M}=(1,2)$ array of unequal SNAILs. The left and right-hand side SNAILs have additional built-in flux focusers leading to the general case where $\mathbf{R}=(1,R_2)$ where in this geometry the middle SNAIL has $R_1=1$ and then edge SNAILs $R_2>1$. (b) Optical micrograph and schematic of the reflection measurement setup. (c) Optical micrograph of an $\mathbf{M}=(1,1,1)$ unequal SNAIL array that is flux-biased differently across the array with flux focusing factors $\mathbf{R}=(1,R_2,R_3)$.}
\label{fig:1}
\end{figure}

An example of the configuration experimentally studied in this work is shown in Fig.~\ref{fig:1} where SNAILs are the JJE and $M=3$. The SNAILs are unequal because flux-focusers (planar superconductors which expel magnetic field by the Meissner effect) embedded in the JPA pads differentially focus flux into the outer SNAILs in the array (Fig.~\ref{fig:1}(a)). This increases their effective area $A_{i}$ so that they are flux-biased differently from SNAILs in the centre of the array. The array is embedded into the centre~$\lambda$/2 superconducting transmission line resonator (Fig.~\ref{fig:1}(b)), and operated in a phase preserving three-wave mixing regime. A micrograph of the SNAIL-array is shown in Fig.~\ref{fig:1}(c) with the built-in flux-focusers labeled. We draw the reader's attention to the different SNAIL-loop terminations at the flux-focusing launching pads. On the left-hand-side leads are inset from the edge of the flux-focuser and on the right-hand-side leads extend out from the top of the flux focuser which will lead to a smaller (larger) flux focusing effect on the leftmost (rightmost) SNAIL. Flux-focusers have been previously used to improve performance in single dc SQUID devices \cite{ketchen1985dc,miklich1991sensitive} as well as dc SQUID arrays \cite{chesca2015flux,chesca2021squid}. Alternative implementations of arrays of unequal JJEs could be realised by modulating loop areas (applying different amounts of flux) or by modulating the Josephson critical current of various Josephson junctions within the array (for different additional implementations see Appendix B). 

The system's resonant frequency has a nonperiodic field dependence and is specifically illustrated in the case of flux-focusing SNAILs which leads to two interesting features. Firstly, (A), that not all JJEs in the array reach specific working points at the same applied bias, for instance the state of optimum gain or Kerr-free flux bias points. Moreover, the different JJEs have different flux dependencies for their individual Josephson critical currents $I_{\mathrm{c},\mathrm{JJE}}(\phi_{\mathrm{ext}})$, and Josephson inductances, where $\phi_{\mathrm{ext}} = \Phi_{\mathrm{ext}}/\Phi_0$ is the flux bias in units of flux quanta. Consequently, this leads to the second feature, (B), that the rf pump, in normalized units $I_{\mathrm{p}}/I_{\mathrm{c}}(\phi_{\mathrm{ext}})$, becomes inhomogeneous across the array. These two features (A) and (B) are central to the self-impedance matching effects which we observe in these devices.

In the following, we define each JJE to have $N-1$ large junctions with nominally identical critical currents $I_{\mathrm{c}}$, while the remaining junction has a smaller critical current $\alpha I_{\mathrm{c}}$, with $0<\alpha<1$. We neglect the geometric inductance $L_{\mathrm{S}}$ of the JJE quantization loops because, in our case, the JJE's screening (or hysteresis) parameter $\beta_{\mathrm{S}} = 2\pi L_{\mathrm{S}}\alpha I_{\mathrm{c}}/\Phi_0 \ll 1$ and therefore its significance in the system's dynamics can be ignored \cite{chesca2015flux,chesca2021squid}. $\Phi_0$ is the magnetic flux quantum. For the case when $\beta_{\rm S}$ is not negligible see Appendix A.

\subsection{The potential and resonant frequency of JPAs based on unequal arrays of Josephson junction elements}

We consider an array of $M=M_1+M_2+\ldots+M_P$ JJEs distinguished by different flux-focusing ratios. The array consists of $P$ different types such that $M_{1}$ JJEs have an effective area $A_1=R_1A$, $M_{2}$ JJEs have an effective area $A_{2}=R_{2}A$ and so on. 
We denote this using vector notation $\boldsymbol{\mathrm{M}} = (M_1, M_2,\ldots, M_P)$ which corresponds to a vector of flux focusing ratios $\boldsymbol{\mathrm{R}} = (R_1, R_2,\ldots, R_P)$.
The superconducting phase differences across the smallest junction in each JJE is denoted by $\phi_{\mathrm{s}i}$. For the case when JJEs are SNAILs see Fig.~\ref{fig:1}(a). Since for arrays of unequal SNAILs the superconducting phase splits unequally across the SNAILs, for every applied flux $\Phi_{\mathrm{ext}}$, we have $\phi_{\mathrm{s}1}\neq\phi_{\mathrm{s}2}\neq\ldots\neq\phi_{\mathrm{s}P}$. The array's $P$-dimensional potential energy can be written as (see also Appendix A):

\begin{subequations}
\label{eq:array-potential}
\begin{equation}
\label{eq:array-potential-sum}
\begin{split}
U_{\mathrm{array}}(\phi_{\mathrm{s}1},\phi_{\mathrm{s}2},\ldots,\phi_{\mathrm{s}P})
&= \sum_{k=1}^{P} M_k U_k(\phi_{\mathrm{s}k}) .
\end{split}
\end{equation}

\begin{equation}
\label{eq:array-potential-element}
\begin{split}
U_k(\phi_{\mathrm{s}k})
&= E_{\mathrm{J},k}\biggl[-\alpha_k\cos\phi_{\mathrm{s}k} \\
&\quad -(N_k-1)\cos\left(
\frac{R_k 2\pi\phi_{\mathrm{ext}}-\phi_{\mathrm{s}k}}{N_k-1}
\right)\biggr],\\
&\hspace{3.0cm} k=1,2,\ldots,P .
\end{split}
\end{equation}
\end{subequations}

where $R_k\phi_{\mathrm{ext}}$ is the normalized externally applied flux to individual JJEs, $E_{\mathrm{J},k} = I_{\mathrm{c},k}\Phi_0/2\pi$, $\alpha_k$ is the asymmetry between the critical current in the two branches of the loop and $N_k$ is the number of Josephson junctions in each JJE which, as mentioned previously, controls whether the JJE is a form of SQUID, SNAIL or other device. For the case when JJEs are rf SQUIDs that have just one junction in the loop $(N_k=1)$, only the first term in Eq.~\ref{eq:array-potential-element} should be considered while  retaining the energy contribution due to a non-negligible $\beta_S$ (see Appendix A). The expression for the potential energy Eq.~\ref{eq:array-potential-element} assumes that $E_{\mathrm{J},k} \gg E_{\mathrm{C_J}}$ for each junction, where $E_{\mathrm{C_J}}=e^2/(2C_{\mathrm{J}})$ is the Coulomb charging energy of the junction with capacitance $C_{\mathrm{J}}$, and when $C_0 \ll C_{\mathrm{J}}/(N_k -1)^2$, where $C_0$ is the capacitance to ground of each island between junctions.

The multi-dimensionality feature in the phase space for the potential of arrays of unequal JJEs is in contrast to the case of arrays of identical JJEs that are described by a one-dimensional potential. This results in quasi-non-periodic flux dependence of the potential in applied flux, $\Phi_{\mathrm{ext}}$, for arrays of unequal JJEs where $\Phi_{\mathrm{ext}}$ is the flux applied to the JJE loop with the smallest effective area, in contrast to the $\Phi_0$ periodic modulation for arrays of identical JJEs.

The total non-linear inductance of the array, $L_{\mathrm{array}}$, is given by the sum of all the generalized Josephson inductances $L_{\mathrm{J},k}$ corresponding to the $M_k$ JJEs, respectively,

\begin{subequations}
\label{eq:array-inductance}
\begin{equation}
\label{eq:array-inductance-total}
L_{\mathrm{array}}\left(\phi_{\mathrm{ext}}\right)
= \sum_{k=1}^{P} M_k L_{\mathrm{J},k}\left(\phi_{\mathrm{ext}}\right)
\end{equation}

\begin{equation}
\label{eq:josephson-inductance}
\frac{1}{L_{\mathrm{J},k}(\phi_{\mathrm{ext}})}=  \left(\frac{2\pi}{\Phi_0}\right)^2 \frac{\mathrm{d}^2U_k(\phi_{\mathrm{s}k})}{\mathrm{d}\phi_{\mathrm{s}k}^2}\qquad k=1,2,\ldots,P 
\end{equation}
\end{subequations}

where JJE inductances are related to the second order partial derivatives of the potential $U_{k}$ given by Eq.~\ref{eq:array-potential}b and calculated using a value of $\phi_{\mathrm{s}k}=\phi_{\mathrm{s}k,min}(\phi_{\rm ext})$, where we have minimised the potential given in Eq.~\ref{eq:array-potential-element}. Unlike the case of identical JJEs arrays with $\Phi_0$ periodic $L_{\mathrm{array}}$, for arrays made of unequal JJEs, $L_{\mathrm{array}}$, like $U_{\mathrm{array}}(\Phi_{\mathrm{ext}})$, is quasi-non periodic with flux, $\Phi_{\mathrm{ext}}$.

Considering the specific realisation of a JPA shown in Fig.~\ref{fig:1}(b) with an array of unequally flux-focused SNAILs ($N_k = 4$) embedded in a $\lambda/2$ resonator we find the total potential energy to be given by:

\begin{equation}
\label{eq:jpa-potential}
\begin{split}
U_{\mathrm{JPA}}(\phi_{\mathrm{s}1},\phi_{\mathrm{s}2},\ldots,\phi_{\mathrm{s}P},\phi)
&= \frac{1}{2\beta}E_{\mathrm{J}}
\left(\phi-\sum_{k=1}^{P}M_k\phi_{\mathrm{s}k}\right)^2 \\
&\quad + U_{\mathrm{array}}(\phi_{\mathrm{s}1},\phi_{\mathrm{s}2},\ldots,\phi_{\mathrm{s}P}) .
\end{split}
\end{equation}

where $\phi$ is the resonator mode canonical phase coordinate, $\beta = 2\pi L I_{\mathrm{c}}/\Phi_0$, $L$ is the linear resonator inductance, $E_{\mathrm{J}} = I_{\mathrm{c}}\Phi_0/2\pi$ and \(U_{\mathrm{array}}\) is given by Eqs.~\ref{eq:array-potential-sum} and \ref{eq:array-potential-element}. In this semi-classical analysis, as the coordinates $\phi_{\mathrm{s}1},\phi_{\mathrm{s}2}$, \ldots, $\phi_{\mathrm{s}P}$ are not independent variables, we can eliminate them by minimizing the potential $U_{\mathrm{JPA}}(\phi_{\mathrm{s}1},\phi_{\mathrm{s}2},\ldots,\phi_{\mathrm{s}P},\phi)$ with respect to $\phi_{\mathrm{s}1},\phi_{\mathrm{s}2},\ldots,\phi_{\mathrm{s}P}$ (see also appendix A). In doing so the problem becomes 1D: \(U_{\mathrm{JPA}} = U_{\mathrm{JPA}}(\phi)\) .

\begin{figure}
\centering
\includegraphics[width=0.5\textwidth]{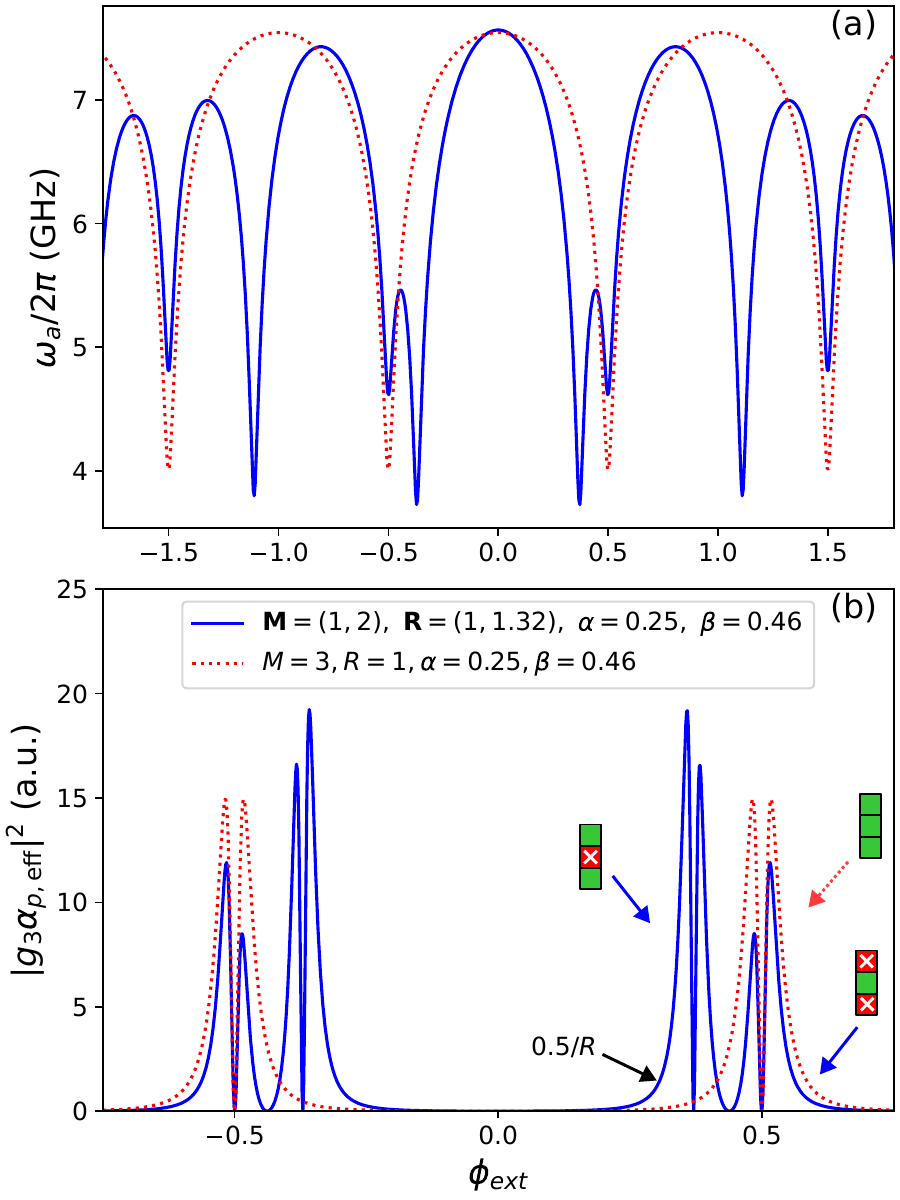}
\caption{Arrays of nominally identical SNAILs versus arrays of
unequal SNAILs. (a) Calculated $\omega_{\mathrm{a}}(\phi_{\mathrm{ext}})/2\pi$ and (b) $|g_3\alpha_{\mathrm{p},\mathrm{eff}}(\phi_{\mathrm{ext}})|^2$ (in arbitrary units a.u.) based on Eqs.~\ref{eq:resonant-frequency} and \ref{eq:effective-pump}, respectively. In both panels solid lines are for a JPA based on an array of unequal SNAILs with $\boldsymbol{\mathrm{M}} = (1,2)$ and $\boldsymbol{\mathrm{R}} = (1, 1.32)$ while dotted lines are for a JPA based on an array of $M=3$ identical SNAILs ($R=1$). In (b) the three-wave mixing active (inactive) SNAILs are schematically represented as green (red with superimposed X) squares. In the calculations the following values for the remaining relevant parameters were used: $\alpha = 0.25$, and $\beta = 0.46$.}
\label{fig:2}
\end{figure}

Approximating the system as a lumped-element resonator we find the resonant frequency of the JPA to be:

\begin{equation}
\label{eq:resonant-frequency}
\omega_{\mathrm{a}}\left(\phi_{\mathrm{ext}}\right)
= \frac{1}{\sqrt{L_{\mathrm{JPA}}C_{\mathrm{JPA}}}}
= \frac{1}{\sqrt{\left(L+L_{\mathrm{array}}(\phi_{\mathrm{ext}})\right)C_{\mathrm{JPA}}}} .
\end{equation}

where the total JPA inductance $L_{\mathrm{JPA}}$ consists of the linear resonator inductance $L$ connected in series with the non-linear array's inductance $L_{\mathrm{JPA}} =L+L_{\mathrm{array}}$ and $C_{\mathrm{JPA}}$ is the equivalent JPA's capacitance. The dependence $\omega_{\mathrm{a}}(\phi_{\mathrm{ext}})$ for a JPA based on an array of unequal SNAILs with $\boldsymbol{\mathrm{M}} = (1,2)$ and $\boldsymbol{\mathrm{R}} = (1, 1.32)$, $\alpha$ =0.25, $\beta$ =0.46 is shown in Fig.~\ref{fig:2}(a). In this case $\omega_{\mathrm{a}}(\phi_{\mathrm{ext}})$ is quasi-non-periodic with flux unlike the case of arrays of identical SNAILs which have a $\Phi_0$ periodic dependence with flux (dotted line in Fig.~\ref{fig:2}(a)).

We operate the JPA in one of its equivalent (equilibrium) minima $\phi_{\mathrm{min}}$ given by
\(\mathrm{d}U_{\mathrm{JPA}}/\mathrm{d}\phi\left( \phi_{\mathrm{min}} \right) = 0\ \) and by Taylor expanding $U_{\mathrm{JPA}}$ given by Eq.~\ref{eq:jpa-potential} around this point we get:

\begin{subequations}
\label{eq:taylor-expansion}
\begin{equation}
\label{eq:taylor-expansion-potential}
\begin{split}
U_{\mathrm{JPA}}(\phi)
&= E_{\mathrm{J}}\biggl[
\frac{c_2}{2!}(\phi-\phi_{\mathrm{min}})^2 \\
&\quad + \frac{c_3}{3!}(\phi-\phi_{\mathrm{min}})^3 + \ldots \biggr]
\end{split}
\end{equation}

where $c_n$ are the Taylor coefficients

\begin{equation}
\label{eq:taylor-coefficients}
c_n = \frac{1}{E_{\mathrm{J}}}\frac{\mathrm{d}^n U_{\mathrm{JPA}}}{\mathrm{d}\phi^n}(\phi_{\mathrm{min}}),\qquad n=1,2,3,\ldots .
\end{equation}
\end{subequations}

calculated in appendix A. As we operate the JPA in one of its potential minima, the first Taylor coefficient is zero: $c_1 = 0$.

\subsection{Gain in JPAs based on unequal arrays of JJEs}

To calculate the gain G of a JPA we analyse the quantum mechanical behavior of the system. We expand the system's Hamiltonian around the harmonic oscillator (linear) approximation truncated to fourth order \cite{frattini2018optimizing,sivak2019kerr}:

\begin{subequations}
\label{eq:jpa-hamiltonian}
\begin{equation}
\label{eq:jpa-hamiltonian-expansion}
\begin{split}
\frac{H_{\mathrm{JPA}}}{\hbar}
&= \omega_{\mathrm{a}} a^\dagger a
+ g_3(a+a^\dagger)^3 \\
&\quad + g_4(a+a^\dagger)^4 .
\end{split}
\end{equation}
where
\begin{equation}
\label{eq:nonlinearity-coefficients}
\begin{split}
g_i
&= \frac{1}{i!}\frac{c_i}{c_2}\frac{\omega_{\mathrm{a}}}{2}
\left(\frac{4E_{\mathrm{C}}}{\hbar\omega_{\mathrm{a}}}\right)^{i/2-1},\\
&\hspace{3.0cm} i=3,4 
\end{split}
\end{equation}
\end{subequations}

where $g_{3}$ and $g_{4}$ are the third- and fourth-order nonlinearity coefficients, respectively. The $g_{3}$ and $g_{4}$ coefficients depend on $\omega_{\mathrm{a}}$ and $c_{n}$ coefficients calculated earlier (see Eqs.~\ref{eq:resonant-frequency} and \ref{eq:taylor-coefficients}, respectively). Here \(a^{\dagger}\) and \(a\) are the bosonic raising and lowering operators respectively and $E_{\rm C} = e^2/2C_{\rm JPA}$.

Using standard input-output theory the intracavity fields \(\widehat{a}\lbrack\omega\rbrack = \left( {\widehat{a}}_{\mathrm{s}}\lbrack\omega\rbrack,{\widehat{a}}_{\mathrm{i}}^{\dagger}\lbrack - \omega\rbrack\ \right)^{T}\) where $\widehat{a}_{\mathrm{s}},~\widehat{a}_{\mathrm{i}}$ are the signal and idler fields respectively \cite{roy2015broadband,frattini2018optimizing,sivak2019kerr,westig2018josephson}, are related to the input fields \({\widehat{a}}_{\mathrm{in}}\lbrack\omega\rbrack = \left( {\widehat{a}}_{\mathrm{in}}\lbrack\omega\rbrack,{\widehat{a}}_{\mathrm{in}}^{\dagger}\lbrack - \omega\rbrack\ \right)^{T}\)via the inverse of the susceptibility matrix ($\chi$), \(\chi^{- 1}\lbrack\omega\rbrack\widehat{a}\lbrack\omega\rbrack = {\widehat{a}}_{\mathrm{in}}\lbrack\omega\rbrack\), which for a JPA operating in a three-wave mixing regime, is given by:

\begin{equation}
\label{eq:susceptibility-matrix}
\begin{gathered}
\chi^{-1}[\omega,\alpha_{\mathrm{p}}]
= -i\begin{pmatrix}
A_{\mathrm{s}} & 4ig_3 \alpha_{\mathrm{p}} \\
4i(g_3 \alpha_{\mathrm{p}})^* & A_{\mathrm{i}}
\end{pmatrix},\\
A_{\mathrm{s}}=\omega-\Delta_{\mathrm{eff}}[\omega]-\Sigma_{\mathrm{s}}[\omega],\\
A_{\mathrm{i}}=\omega+\Delta_{\mathrm{eff}}[\omega]-\Sigma_{\mathrm{i}}[\omega].
\end{gathered}
\end{equation}

Here $\alpha_{\mathrm{p}}$ is the classical mean intracavity pump amplitude controlled by the input normalized rf pump amplitude $I_{\mathrm{pn}}$ = $I_{\mathrm{p}}/I_{\mathrm{c}}$ at the JPA, \(\omega = \omega_{\mathrm{s}} - \omega_{\mathrm{p}}/2\ \) is the signal detuning from the pump tone, \(\Delta_{\mathrm{eff}} = \omega_{\mathrm{a}} - \omega_{\mathrm{p}}/2\) is the effective pump detuning that depends on the pumping strength and \(\Sigma_{\mathrm{s},\mathrm{i}}\) are the self-energies \cite{roy2015broadband} related to the input impedance seen by the JPA at the signal (subscript s) frequency $+\omega$ and idler (subscript i) frequency $-\omega$, respectively. For a frequency-independent environment when parametric amplifiers are connected to output lines having a purely ohmic impedance $R_{0}=50$~$\Omega$, \(\Sigma_{\mathrm{s},\mathrm{i}}\) does not depend on $\omega$ \cite{frattini2018optimizing,sivak2019kerr}. In contrast, when impedance matching circuits are implemented in order to increase the instantaneous bandwidth of the amplifier, the self-energies are both frequency dependent: \(\Sigma_{\mathrm{s},\mathrm{i}} = \Sigma_{\mathrm{s},\mathrm{i}}(\omega)\) \cite{roy2015broadband}.

The product of the off-diagonal terms in the matrix given by Eq.~\ref{eq:susceptibility-matrix} is proportional to $\left| g_{3}\alpha_{\mathrm{p}} \right|^{2}$ which is a measure of the effective three-wave mixing interaction \cite{roy2015broadband,frattini2018optimizing}. For optimal performance, JPAs are operated in the linear regime, where $\alpha_{\mathrm{p}}$ is proportional to $I_{\mathrm{pn}}$. For JPAs based on arrays of identical JJEs, $I_{\mathrm{pn}}$ and therefore $\alpha_{\mathrm{p}}$ are both homogeneous across the array. For JPAs based on arrays of unequal JJEs the situation is different. If we consider the case where JJEs are distinguished by different rates of applied flux, the flux dependencies of the Josephson critical current $I_{\mathrm{c},\mathrm{JJE}}(\phi_{\mathrm{ext}})$ of JJEs with higher applied flux  are shifted by $(R-1)/R$ in normalised flux resulting in different instantaneous values of $I_{\mathrm{c}}$. As a result, the rf pump in normalized units $I_{\mathrm{pn}}$=$I_{\mathrm{p}}/I_{\mathrm{c}}$ is inhomogeneous across the array meaning that for most flux biases the JJEs are pumped differently unlike the situation with identical JJEs. 
In other words, although the same rf current flows through the series-connected Josephson elements, their different flux-dependent inductance produces unequal phase responses. The resulting spatially inhomogeneous phase amplitudes modify the effective nonlinear interaction.
Consequently, in amplifiers based on arrays made of unequal JJEs, the gain is determined by \(\left| g_{3}\alpha_{\mathrm{p},\mathrm{eff}} \right|^{2}\ \) rather than \(\left| g_{3}\alpha_{\mathrm{p}} \right|^{2}\), where $\alpha_{\mathrm{p},\mathrm{eff}}$ is the effective intracavity pump amplitude that takes into account the inhomogeneous rf pumping across the array:

\begin{equation}
\label{eq:effective-pump}
\begin{split}
\left|g_3 \alpha_{\mathrm{p},\mathrm{eff}}\right|^2
&= \left|g_3\alpha_{\mathrm{p}}\frac{L_{\mathrm{array}}}{L_{\mathrm{J}}}\right|^2 \\
&= \left(\frac{1}{3!}\alpha_{\mathrm{p}}\frac{1}{L_{\mathrm{J}}}\right)^2
\frac{E_{\mathrm{C}}}{\hbar}
\left|\frac{c_3}{c_2}\sqrt{\omega_{\mathrm{a}}} L_{\mathrm{array}}\right|^2 .
\end{split}
\end{equation}

All four terms inside the modulus of Eq.~\ref{eq:effective-pump} have an explicit dependence on $\phi_{\mathrm{ext}}$. Therefore, to calculate
\(\left| g_{3}\alpha_{\mathrm{p},\mathrm{eff}} \right|^{2}\left( \phi_{\mathrm{ext}} \right)\)
using Eq.~\ref{eq:effective-pump} we need the explicit $\phi_{\mathrm{ext}}$ dependencies of $c_{2}$ and $c_{3}$ (see Eqs.~\ref{eq:appendix-c2-coefficient_general} and \ref{eq:appendix-c3-coefficient_general} in appendix A, respectively), as well as $\omega_{\mathrm{a}}(\phi_{\mathrm{ext}})$ given by Eq.~\ref{eq:resonant-frequency} and $L_{\mathrm{array}}(\phi_{\mathrm{ext}})$ given by Eqs.~\ref{eq:array-inductance-total} and \ref{eq:josephson-inductance}. \(\left| g_{3}\alpha_{\mathrm{p},\mathrm{eff}} \right|^{2}\left( \phi_{\mathrm{ext}} \right)\) dependencies are important since they provide us with the optimal flux bias point of the amplifier where gain, signal to noise improvement and instantaneous bandwidth are optimized. To conclude, for JPAs based on arrays of unequal SNAILs, the susceptibility matrix given by Eq.~\ref{eq:susceptibility-matrix} should be rewritten with the transformation: $\alpha_{\mathrm{p}} \rightarrow \alpha_{\mathrm{p},\mathrm{eff}}$.

In Fig.~\ref{fig:2}(b) we highlight the difference in operation between JPAs containing three identical SNAILs and an array of unequal SNAILs plotting $|g_3\alpha_{\mathrm{p},\mathrm{eff}}|^2$, a measure of the effective three-wave mixing interaction in these devices, as a function of applied flux. When SNAILs are identical, three-wave mixing gain turns on at the same flux for all SNAILs resulting in periodic behaviour with two equal peaks per $\Phi_0$ period. In Fig.~\ref{fig:2}~(b) we indicate these regions where the SNAILs collectively `turn on' with a cartoon of three green squares indicating that all three SNAILs are three-wave active simultaneously.

This contrasts with the calculation of $|g_3\alpha_{\mathrm{p},\mathrm{eff}}|^2$ performed for a JPA based on an array of unequal SNAILs distinguished by degree of flux-focusing with $\boldsymbol{\mathrm{M}} = (1,2)$, $\boldsymbol{\mathrm{R}} = (1,1.32)$, $\alpha=0.25$ and $\beta$ = 0.46. For this device, there is no periodicity and the peaks have different amplitudes. The peaks are distinguished both by (i) the SNAILs which are three-wave active at a given external flux-bias (ii) the rf pump amplitude across the SNAILs determined by the uneven phase drop across SNAIL elements. The cartoon representations of these working points have a mix of green and red squares, showing that not all SNAILs in the array are three-wave active. Comparing the cartoons of three-wave active amplifiers for identical vs. unequal SNAILs, it is apparent that in the case of unequal SNAILs, peaks in gain occur when a subset of SNAILs have large $|g_3\alpha_{\mathrm{p},\mathrm{eff}}|^2$, and other SNAILs are contributing only a finite lumped inductance. We highlight here that this cartoon (which is repeated in additional figures) is a helpful illustration but in reality three-wave mixing in SNAILs will turn on/off by degree, something captured in our full calculations.

Let us now consider the case when an $LC$ resonator (inductance $L_{\mathrm{M}}$ and capacitance $C_{\mathrm{M}}$) is used as an impedance matching circuit for the purpose of increasing the amplifier instantaneous bandwidth. In our circuits this would be realised by the tunable inductance of the three-wave inactive JJEs. The self-energies \(\Sigma_{\mathrm{s},\mathrm{i}}\) have the following explicit frequency dependencies \cite{roy2015broadband}:

\begin{subequations}
\label{eq:self-energies}
\begin{equation}
\label{eq:self-energy-definition}
\Sigma_{\mathrm{s},\mathrm{i}}[\omega]=\Delta_{\mathrm{s},\mathrm{i}}[\omega]-i\kappa_{\mathrm{s},\mathrm{i}}[\omega]/2
\end{equation}

where for small $\omega$ detuning one can approximate

\begin{equation}
\label{eq:self-energy-detuning}
\Delta_{\mathrm{s},\mathrm{i}}[\omega]
= \frac{1}{C_{\mathrm{p}}}\frac{\omega L_{\mathrm{M}}}{4(\omega L_{\mathrm{M}})^2+R_{0}^2} .
\end{equation}

\begin{equation}
\label{eq:self-energy-kappa}
\kappa_{\mathrm{s},\mathrm{i}}[\omega]
= \frac{R_{0}}{C_{\mathrm{p}}}\frac{4}{4(\omega L_{\mathrm{M}})^2+R_{0}^2} .
\end{equation}
\end{subequations}

Here $C_{\mathrm{p}}$ is the shunting capacitance of the JJE array and $R_{0}=50~\Omega$ is the environment resistance. At small pump detuning these equations are independent of $C_{\mathrm{M}}$. By inverting \(\chi^{- 1}\) given by Eq.~\ref{eq:susceptibility-matrix} and evaluating $\chi_{11}$ after replacing $\alpha_{\mathrm{p}}$ with $\alpha_{\mathrm{p},\mathrm{eff}}$, to account for an array of unequal SNAILs, the gain can be calculated as:

\begin{subequations}
\label{eq:gain-profile}
\begin{equation}
\label{eq:gain}
\begin{split}
G[\omega,\alpha_{\mathrm{p},\mathrm{eff}}]
&= \left|1-\kappa_{\mathrm{s}}[\omega]\chi_{11}[\omega]\right|^2 \\
&= \left|1-\kappa_{\mathrm{s}}[\omega]
\frac{i\left(\omega+\Delta_{\mathrm{eff}}[\omega]-\Sigma_{\mathrm{i}}[\omega]\right)}
{D[\omega,\alpha_{\mathrm{p},\mathrm{eff}}]}
\right|^2 .
\end{split}
\end{equation}
where
\begin{equation}
\label{eq:gain-determinant}
\begin{split}
D[\omega,\alpha_{\mathrm{p},\mathrm{eff}}]
&= \left(\omega-\Delta_{\mathrm{eff}}[\omega]-\Sigma_{\mathrm{s}}[\omega]\right) \\
&\quad \times \left(\omega+\Delta_{\mathrm{eff}}[\omega]-\Sigma_{\mathrm{i}}[\omega]\right)
+16\left|g_3 \alpha_{\mathrm{p},\mathrm{eff}}\right|^2 .
\end{split}
\end{equation}
\end{subequations}

Eqs.~\ref{eq:gain} and \ref{eq:gain-determinant} can now be used to calculate the gain profile $G[\omega]$.

\begin{figure}
\centering
\includegraphics[width=0.5\textwidth]{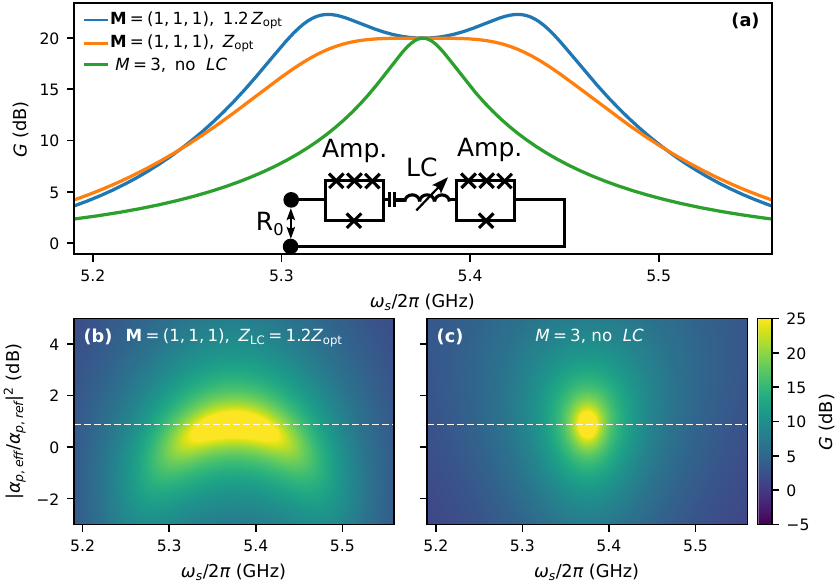}
\caption{Calculated $G(\omega)$ using Eqs.~\ref{eq:gain} and \ref{eq:gain-determinant} for two cases. The first case is for an array of unequal SNAILs with $\boldsymbol{\mathrm{M}} = (1,1,1)$, $\boldsymbol{\mathrm{R}} = (1,1.32,1.40)$ and modelled using an effective circuit shown in the inset of (a): the first and third SNAILs in the array operate as tuned parametric amplifiers, while the middle SNAIL operates as an LC resonator with a tunable inductance $L_{\mathrm{M}}$, or equivalently, tunable impedance $Z_{\mathrm{LC}}$; in (a) the blue curve is for $Z_{\mathrm{LC}}=1.2 Z_{\mathrm{opt}}$ and the orange curve is for $Z_{\mathrm{LC}}= Z_{\mathrm{opt}}$, while the heatmap in (b) is for $Z_{\mathrm{LC}}=1.2 Z_{\mathrm{opt}}$. The second case is for an array of $M=3$ nominally identical SNAILs (i.e., $R=1$) and no LC resonator: see the green curve in (a) and heatmap in (c). (a) $G(\omega)$ for a pump power corresponding to the horizontal dotted lines in (b) and (c), and a pump frequency of $2\times5.375$ GHz. Gain is calculated for different effective intracavity pump amplitude $|\alpha_{\mathrm{p},\mathrm{eff}}/\alpha_{\mathrm{p},\mathrm{ref}}|^2$. We normalize $\alpha_{\mathrm{p},\mathrm{eff}}$ to $\alpha_{\mathrm{p},\mathrm{ref}}$ which is the intracavity pump amplitude required to produce 20 dB central gain.}
\label{fig:3}
\end{figure}

Here we demonstrate how we can build an LC matching circuit that has a variable impedance due to its flux-tunable Josephson inductance $L_{\mathrm{J}}$ and a small footprint. Its functionality is qualitatively similar to purpose built passive, fixed-frequency matching circuits \cite{roy2015broadband,mutus2014strong,ezenkova2022broadband,kaufman2023josephson,white2023readout,joshi2025lumped,westig2018josephson,getsinger1963prototypes,naaman2022synthesis}. We consider the case of an array of unequal SNAILs with $\boldsymbol{\mathrm{M}} = (1,1,1)$ replicating the flux-focusing array created with SNAIL layouts such as those shown in Fig.~\ref{fig:1}(c). As Fig.~\ref{fig:2}(b) suggests the different working points of the SNAILs means that they take on different functions in the microwave circuit. When biased away from favourable three-wave mixing points the value of $|g_{3}\alpha_{\mathrm{p},\mathrm{eff}}|^{2}$is small, and the SNAIL is effectively \textquotesingle turned off\textquotesingle{} meaning they contribute negligibly to wave-mixing which ultimately is what results in gain. They do, however, still have an inductance which, together with the network capacitance, form an LC circuit which modifies the microwave properties of the device. Those SNAILs operate as flux tunable matching circuits rather than parametric amplifiers. When biased appropriately they \textquotesingle turn on\textquotesingle{} with a large value for $|g_{3}\alpha_{\mathrm{p},\mathrm{eff}}|^{2}$ and still some finite inductance. Let us now consider a particular flux bias point where the first and third SNAILs operate as tunable parametric amplifiers while the middle SNAIL operates as a tunable Josephson inductance (or tunable impedance $Z_{\mathrm{LC}} = \sqrt{L_{\mathrm{M}}/C_{\mathrm{M}}}$) LC resonator (see inset of Fig.~\ref{fig:3}(a) for the circuit used to model these devices). We use Eqs.~\ref{eq:gain} and \ref{eq:gain-determinant} to calculate $G(\omega)$ while changing $Z_{\mathrm{LC}}$ and keeping a 20 dB gain at the central frequency (see Fig.~\ref{fig:3}(a)). As we tune $Z_{\mathrm{LC}}$ the width of the instantaneous bandwidth has a highly non-monotonic dependence with high values reached in a rather narrow range around the optimum impedance for flat-band gain $Z_{\mathrm{opt}}$. For $Z_{\mathrm{LC}}$ around $Z_{\mathrm{opt}}$, the gain profile for values above 20 dB changes from a double-peaked gain profile (blue curve in Fig.~\ref{fig:3}(a)) to a flat-top gain profile (orange curve in Fig.~\ref{fig:3}(a)) having a bandwidth between 145 and 160 MHz, i.e., about 5 times larger at 17~dB gain than that of a JPA with no $LC$ resonator that shows a Lorentzian, single-peak gain profile with a bandwidth of about 32 MHz (green curve in Fig.~\ref{fig:3}(a)). While the $G(\omega)$ data shown in Fig.~\ref{fig:3}(a) corresponds to fixed pump power values shown with horizontal dotted lines in Figs.~\ref{fig:3}(b, c), a similar increase in the instantaneous bandwidth is predicted for all pump powers where gain above 15 dB is observed. In these calculations we used $Z_{\rm opt} = 506.6~\Omega$, $L_{\mathrm{M}}$~=~15~nH, $C_{\mathrm{M}}$~=~58~fF and $C_{\mathrm{p}}$~=~8.9~pF. We note that in these flux-focused devices, the flux focusing ratio is an important design feature as the impedance matching SNAILs must be tuned to the correct inductance at the same external bias which turns on three-wave mixing in the amplifying SNAILs.

In summary, the three main results for devices with arrays of unequal JJEs from these theory sections are: the quasi-non-periodic flux dependence of the JPA resonant frequency $\omega_{\mathrm{a}}(\phi_{\mathrm{ext}})$ given by Eq.~\ref{eq:resonant-frequency} (see Fig.~\ref{fig:2}(a)), the quasi-non-periodic flux dependence of $|g_3 \alpha_{\mathrm{p},\mathrm{eff}}|^2(\phi_{\mathrm{ext}})$ that is a measure of the effective three-wave mixing interaction, given by Eq.~\ref{eq:effective-pump} (see Fig.~\ref{fig:2}(b)), and the gain versus signal frequency $G[\omega]$ given by Eqs.~\ref{eq:gain} and \ref{eq:gain-determinant} (see Fig.~\ref{fig:3}). In the following section those calculations will be directly compared with measurements performed on JPAs based on two types of arrays of unequal SNAILs characterized by: (a) $P=3$, $\boldsymbol{\mathrm{M}} = (1,1,1)$  and (b) $P=3$, $\boldsymbol{\mathrm{M}} = (3,1,1)$.

\section{JPAs based on arrays of unequal SNAILs: experiments}

We experimentally study the device shown in Fig.~\ref{fig:1} and explored analytically in Fig.~\ref{fig:3}. 
For all devices in this work, we held the microstrip width constant at 437 $\mu$m and have adjusted its length (in conjunction with $M$ and $I_{\mathrm{c}}$) to set the operating frequencies of the devices. The coupling to the 50 $\Omega$ transmission line is set by an interdigitated capacitor at the input of the JPA resonator (left side of the optical image in Fig.~\ref{fig:1}(b)). The JPA devices also have a second, more weakly capacitively coupled, port on the opposite end of the resonator for the delivery of the pump (shown schematically in Fig.~\ref{fig:1}(b)). The experimental characterization was performed in a dilution refrigerator at a temperature of 20~mK with a standard microwave reflection measurement setup. A dc current $I_{\mathrm{B}}$ through a solenoid mounted above the sample generates a magnetic field for flux biasing. All measurements were performed with a standard vector network analyser, which contains two microwave sources, for the pump and signal. The strong pump tone required for amplification was applied on a separate pump line (see Fig.~\ref{fig:1}(b)).

The values of $\alpha$, $\beta$ and $R$ were determined by fitting the measured resonant frequency versus flux curve $\omega_{\mathrm{a}}(\Phi_{\mathrm{ext}})$. All SNAILs have the same estimated geometric inductance $L_{\mathrm{S}}$=68 pH determined by assuming a characteristic length inductance of 1 pH per micrometre. $L$ has been calculated using commercially available software (Electromagnetic Interference Software -- EMI Software LLC). The values for $I_{\mathrm{c}}$ were determined using the formula $\beta =2\pi L I_{\mathrm{c}}/\Phi_0$.

\subsection{JPA frequency dependence on the applied flux}

\begin{figure}
\centering
\includegraphics[width=0.5\textwidth]{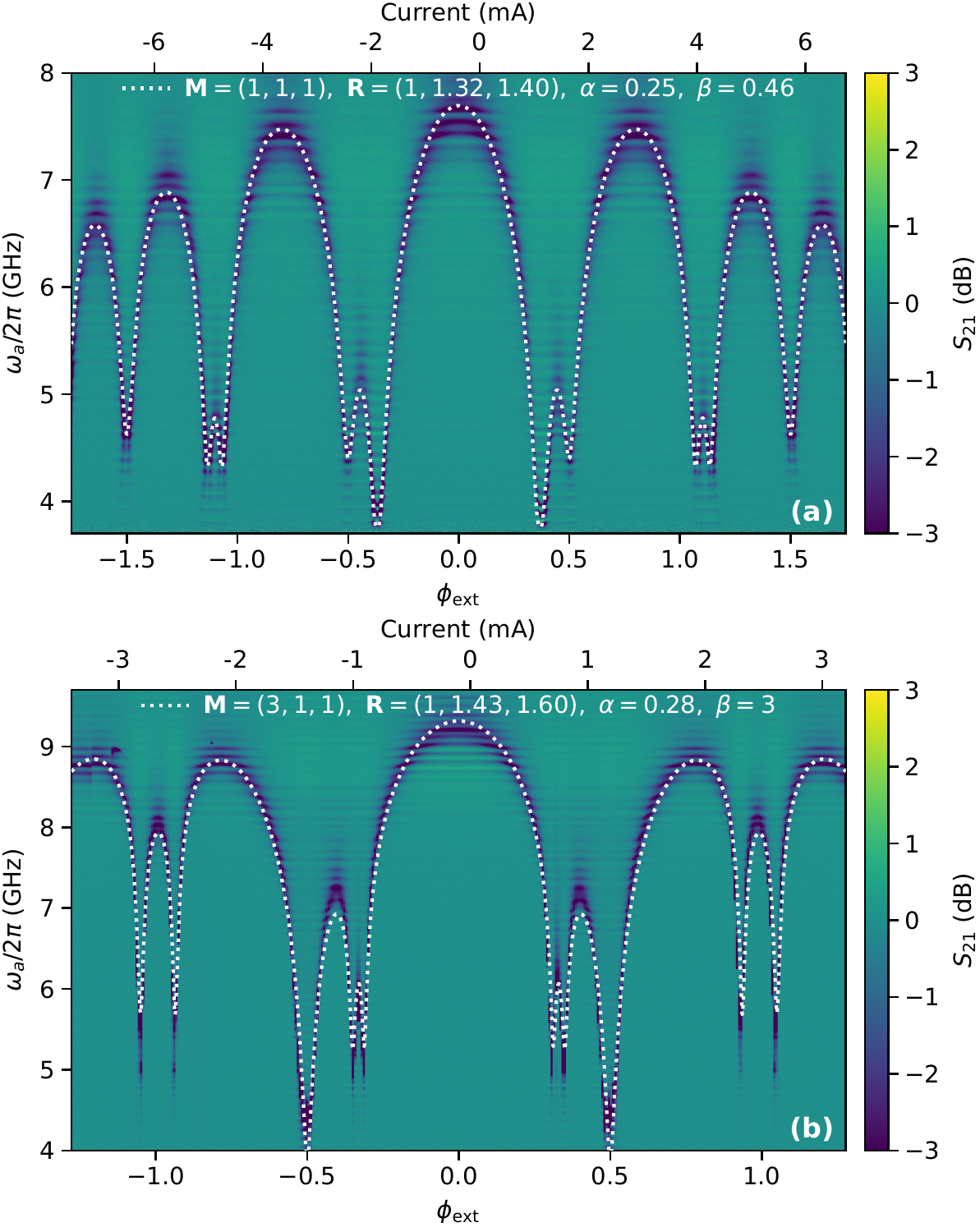}
\caption{Low-power, single-tone measurements of resonant frequency $\omega_{\mathrm{a}}$ versus normalized
externally applied dc magnetic flux $\phi_{\mathrm{ext}}$ of two JPAs based on arrays of unequal SNAILs. In
blue are the absorption single-tone transmission $S_{21}$
heatmap measurements versus the corresponding B field coil current;
white dotted lines are calculations of
$\omega_{\mathrm{a}}(\phi_{\mathrm{ext}})$ based on Eq.~\ref{eq:resonant-frequency} with the following parameters: (a) a device with
$\boldsymbol{\mathrm{M}} = (1,1,1)$,$\boldsymbol{\mathrm{R}} = (1,1.32,1.40)$, $\alpha = 0.25$, $\beta=0.46$;
(b) a device with $\boldsymbol{\mathrm{M}} = (3,1,1)$, $\boldsymbol{\mathrm{R}} = (1,1.43,1.60)$,
$\alpha$=0.28, $\beta$=3. In the measurements for each fixed coil current, the
signal frequency is swept over several GHz.}
\label{fig:4}
\end{figure}

In Fig.~\ref{fig:4} we show single-tone spectroscopy measurements for the device with $\boldsymbol{\mathrm{M}} = (1,1,1)$ (Fig.~\ref{fig:4}(a)) and for the device $\boldsymbol{\mathrm{M}} = (3,1,1)$ (Fig.~\ref{fig:4}(b)) with calculations of $\omega_{\mathrm{a}}/2\pi(\phi_{\mathrm{ext}})$ dependence overlaid. To match experimental data with calculated frequency modulation we tuned $C_{\rm JPA}, \alpha, \beta, R$ to find the best agreement. 
In the fitting procedures for both designs we chose the same value for $C_{\mathrm{JPA}}$ to simulate the capacitive coupling in both cases. As expected, the measured $\omega_{\mathrm{a}}(\Phi_{\mathrm{ext}})$ is symmetrical about an axis that is shifted away from $I_{\mathrm{B}} = 0$ due to a residual magnetic field/flux inside the cryostat. The resonant frequency, $\omega_{\mathrm{a}}(\Phi_{\mathrm{ext}})$, shows no periodicity with $\Phi_{\mathrm{ext}}$ over the measured range and is characterized by a central peak that has a larger modulation depth and width than the subsequent peaks.

\subsection{Gain in JPAs based on unequal SNAIL arrays}

\begin{figure}
\centering
\includegraphics[width=0.5\textwidth]{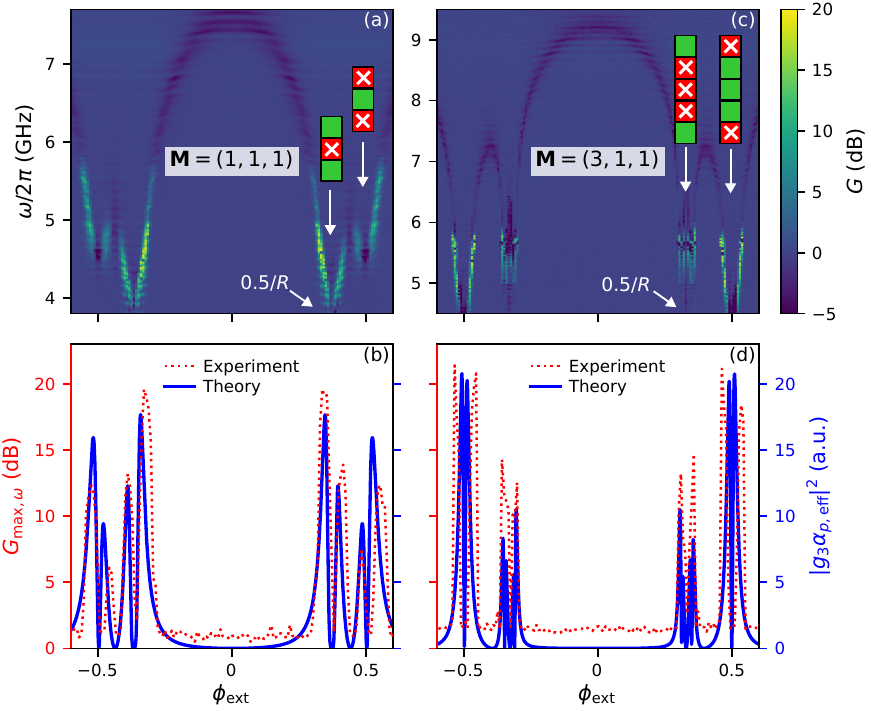}
\caption{(a, c) measured 2D heatmaps for gain $G(\phi_{\mathrm{ext}}, \omega)$, with (b, d) extracted $G_{\mathrm{max},\omega}(\phi_{\mathrm{ext}})$ and calculated $|g_3 \alpha_{\mathrm{p},\mathrm{eff}}|^2(\phi_{\mathrm{ext}})$ based on Eq.~\ref{eq:effective-pump} in arbitrary units (a.u.) for an array of unequal SNAILs with (a) $\boldsymbol{\mathrm{M}} = (1,1,1)$ and (c) $\boldsymbol{\mathrm{M}} = (3,1,1)$. The three-wave mixing active (inactive) SNAILs are schematically represented as green (red with superimposed X) squares, while the vertical arrows indicate the corresponding flux bias regions of applicability. In the calculations the following values for the relevant parameters were used: (b) $\boldsymbol{\mathrm{R}} = (1,1.32,1.40)$, $\alpha$=0.25, and $\beta$=0.46 and (d) $\boldsymbol{\mathrm{R}} = (1,1.43,1.60)$, $\alpha$=0.28 and $\beta$=3. These are precisely the same values used in the theory plots of Fig.~\ref{fig:4}.}
\label{fig:5}
\end{figure}

We operated the amplifiers in the three-wave mixing regime and measured the gain as a function of $\phi_{\mathrm{ext}}$ and frequency for a fixed pump power and a 10 MHz signal detuning from $\omega_{\mathrm{p}}/2$. We measured $S_{21}$ with both pump-on and pump-off and then subtracted pump-off from pump-on to obtain the gain in dB.
In Fig.~\ref{fig:5}(a) and (c), we show representative plots of low-pump-power gain as a function of frequency and applied flux for two JPAs based on arrays of unequal SNAILs with $\boldsymbol{\mathrm{M}} = (1,1,1)$ and $\boldsymbol{\mathrm{M}} = (3,1,1)$ respectively. From these heatmaps we have extracted the maximum gain at each value of flux bias, $G_{\mathrm{max},\omega}(\phi_{\mathrm{ext}})$ and found a good qualitative agreement with the analytical calculations of $|g_3 \alpha_{\mathrm{p},\mathrm{eff}}|^2(\phi_{\mathrm{ext}})$ based on Eq.~\ref{eq:effective-pump} shown in Figs.~\ref{fig:5}(b, d).
There is good qualitative agreement between measurements and calculations despite these calculations being performed using circuit parameters extracted from the frequency modulation of the JPA (Fig.~\ref{fig:4}) with no additional fit parameters.

To measure the JPA operating bandwidth we measure gain and noise-visibility ratio (NVR) by performing four different measurements. Two of them are $S_{21}$ transmission-like measurements required to define gain $G=S_{21}$($S_{\mathrm{on}},P_{\mathrm{on}}) - S_{21}$($S_{\mathrm{on}},P_{\mathrm{off}})$, where $S_{21}$($S_{\mathrm{on}},P_{\mathrm{on}}$) is measured with signal on and pump on while $S_{21}$($S_{\mathrm{on}},P_{\mathrm{off}}$) is performed with signal on and pump off. The other two are required to define NVR = $N$($S_{\mathrm{off}}, P_{\mathrm{on}}$) - $N$($S_{\mathrm{off}}, P_{\mathrm{off}}$) where N($S_{\mathrm{off}}, P_{\mathrm{on}}$) corresponds to signal off and pump on while N($S_{\mathrm{off}},P_{\mathrm{off}}$) to signal off and pump off. Gain and noise values are expressed in dB.
We measured NVR for the $\boldsymbol{\mathrm{M}} = (1,1,1)$ to provide a consistency check on the JPA's noise performance since being in the low noise regime is important for amplifier applications~\cite{joshi2025lumped}. We find the NVR in Fig.~\ref{fig:6}(a) is similar in magnitude to that of devices shown in the appendix of Ref.~\cite{joshi2025lumped} and has bandwidth aligned to the measurement of gain also presented in Fig.~\ref{fig:6}(a). NVR measurements have been performed for all gain curves presented for this device in Fig.~\ref{fig:6}(c) to ensure that these are also not collected from over-driven JPAs. 

For JPAs based on single ($M=1$) SNAILs we measured a $G(\omega)$ that has a Lorentzian shaped profile, characterized by an instantaneous bandwidth of below 33~MHz at 17~dB gain, consistent with previously published reports by other groups on JPAs based on a single SNAIL \cite{frattini2018optimizing}. Relative to this, the amplifiers based on unequal SNAIL arrays show a significant improvement in instantaneous bandwidth, with an example shown in Fig.~\ref{fig:6}(a). This is an increase by a factor between 3-6, up to 200 MHz at 17 dB gain which is also reflected in the G-NVR measurements.  In Fig.~\ref{fig:6}(a) we overlay the analytically derived curve with $Z_{\rm LC} = 1.2 Z_{\rm opt}$ to directly compare our theoretical results to experimental data. The roll off in gain matches between experimental and analytical results although details of the gain profile are not recreated. 
The improvement in bandwidth is relatively stable with flux bias. As an example, for one JPA based on an $\boldsymbol{\mathrm{M}} = (1,1,1)$ array we could flux-tune an instantaneous bandwidth greater than 100 MHz at 17~dB gain across a broad frequency range of 1.5 GHz, (4.9-6.4) GHz as shown in Fig.~\ref{fig:6}(c). The improvement in bandwidth is consistent with the theoretical predictions shown in Fig.~\ref{fig:3}. 
In Fig.~\ref{fig:6}(c) we also present a gain curve from a five-SNAIL device with $\boldsymbol{\mathrm{M}} = (3,1,1)$ which achieves a bandwidth of $\sim$100~MHz. This bandwidth exceeds that measured in JPAs lacking an impedance matcher by us and other groups. This is further evidence that this effect is relatively robust to the details of implementation as here we demonstrate this bandwidth improvement in a device with five SNAILs as well as the three-SNAIL device discussed in detail. 

The qualitative agreement between experiment and calculation suggests that the analytical model captures the main device physics of impedance matching. The largest fractional bandwidth (bandwidth relative to operating frequency) measured in these devices is 4\%, less than the $\sim$10\% representative state-of-the-art bandwidths demonstrated in purpose-built impedance matchers~\cite{mutus2014strong, roy2015broadband, moskaleva2024lumped}, promising for a first generation device.
The good qualitative agreement between the calculated (Fig.~\ref{fig:3}(b)) and measured (Fig.~\ref{fig:6}(b)) gain profiles  allows us to give estimates on the pump power reaching the device, i.e. estimating the pump coupling efficiency. Calculations of internal energy scales indicate -88 dBm pump power at the device plane. Given -68~dB of line attenuation (50~dB of attenuators and $\sim$18~dB estimated line loss from room temperature calibrations), and -11.5~dBm pump power we can obtain an order of magnitude estimate on the pump coupling efficiency of - 88 + 11.5 + 68 $\sim$ -8.5~dB.

\begin{figure}
\centering
\includegraphics[width=0.5\textwidth]{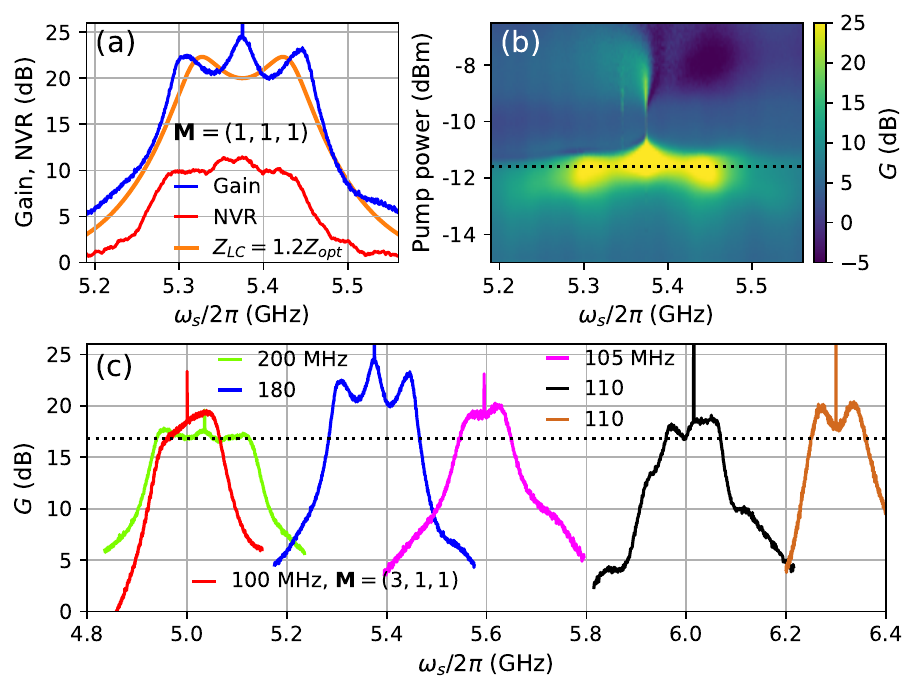}
\caption{Gain measurements of JPAs with arrays of unequal SNAILs. (a) Experimental measurements (blue) of gain (pump on - pump off), (red) NVR with (orange) a theoretical curve calculated for this operating point with SNAILs acting as lumped-element inductors combined with three-wave active amplifying SNAILs. (b) Two-dimensional gain map as a function of signal frequency and pump power, showing the pump-power range over which broadband gain is maintained. (c) Gain curves from the amplifier at different flux bias and consequent frequency working points with their bandwidth at 17~dB gain indicated in the legend. Included on the same axis is a gain curve from the five SNAIL device with $\boldsymbol{\mathrm{M}} = (3,1,1)$ indicated by the legend.}
\label{fig:6}
\end{figure}

\section{Conclusions}

In this work we introduced JPAs based on arrays of unequal JJEs and present a generalised analytical framework to describe these devices. We use the analytical framework to analyse the behaviour of specific devices based on arrays of unequal SNAILs and show that these devices can realise `self impedance matching'. This occurs when some SNAILs in an array act as lumped inductors forming an impedance matching circuit and increasing amplifier bandwidths while other SNAILs are `turned on' with large wave-mixing coefficients. We demonstrate this concept experimentally, operating amplifiers in the three-wave mixing regime. We choose simple implementations of the concept where JPAs are based on arrays of three different types of SNAILs with $\boldsymbol{\mathrm{M}} = (1,1,1)$ and $\boldsymbol{\mathrm{M}} = (3,1,1)$. Our devices show significant improvements in the instantaneous bandwidth relative to JPAs based on series arrays of nominally identical JJEs, such as rf/dc SQUIDs or SNAILs. Our concept is relatively easy to implement in practice and uses the same circuit elements that induce non-linearity into Josephson junction based parametric amplifiers. Similarly to the case of parametric amplifiers based on arrays of nominally identical elements~\cite{kaufman2023josephson}, to increase the amplifier saturation power and improve gain compression larger arrays of JJEs may be implemented.  The concept is transferable to other Josephson junction-based arrays, such as rf-SQUIDs, dc-SQUIDs, three junction SNAILs or any other variation of multi-junction loop arrays operating in either three-wave or four-wave mixing regimes. The promising device figures of merit from our first-generation demonstrations suggest that further optimisation could provide a route to small-footprint, performant amplifiers. In principle with modest modifications to device geometries, i.e., adding a per-SNAIL flux bias line, these devices could allow tunable impedance-matching circuits modulating or optimising amplifier bandwidths.

\appendix
\setcounter{figure}{0}
\renewcommand{\thefigure}{A\arabic{figure}}

\section{Theory of JPAs based on unequal JJE arrays}

In this Appendix we derive a more general equation for the potential without neglecting the geometric inductance $L_{\mathrm{S}}$ of the JJE quantization loops, corresponding to the case where the JJE screening or hysteresis parameter $\beta_{\mathrm{S}}=2\pi L_{\mathrm{S}}\alpha I_{\mathrm{c}}/\Phi_0$ is not negligibly small. When $\beta_{\mathrm{S}}=2\pi L_{\mathrm{S}}\alpha I_{\mathrm{c}}/\Phi_0\ll1$, the generalized potential reduces to the one given by Eqs.~\ref{eq:array-potential-sum} and \ref{eq:array-potential-element}. For this particular case we derive the explicit expressions for the coefficients $c_n$ given by Eq.~\ref{eq:taylor-expansion-potential}.

Here we present a full derivation for the convenience of the reader. We repeat equations~\ref{eq:array-potential-element} and \ref{eq:jpa-potential} from the main text within the derivation for convenience of the reader. Let us start with the potential $U$ of a single JJE within an array of JJEs. Again, we assume that each JJE has $N-1$ large junctions with nominally identical Josephson critical currents $I_{\mathrm{c}}$ and superconducting phase differences $\phi_{\mathrm{large}}$ across each of them, while the remaining junction has a smaller critical current $\alpha I_{\mathrm{c}}$, with $0<\alpha<1$, and a superconducting phase difference $\phi_{\mathrm{s}}$ across it. Using the approach from Ref.~\cite{chesca2004squid} for devices consisting of Josephson junctions in a quantization loop, such as rf and dc SQUIDs, one can write the potential describing each JJE in the array as:

\begin{equation}
\label{eq:appendix-single-jje-potential-general}
\begin{split}
U(j_{\mathrm{n}},\phi_{\mathrm{s}},\phi_{\mathrm{large}})
&=\frac{1}{2}\alpha\beta_{\mathrm{S}} E_{\mathrm{J}} j_{\mathrm{n}}^2 \\
&\quad + E_{\mathrm{J}}\left[-\alpha\cos\phi_{\mathrm{s}}-(N-1)\cos\phi_{\mathrm{large}}\right].
\end{split}
\end{equation}

Here $j_{\mathrm{n}}=J/(\alpha I_{\mathrm{c}})$ is the normalized circulating current in the JJE loop. We use this particular normalization to emphasize that $J$ can never exceed the value of the smallest critical current in the JJE loop, so $j_{\mathrm{n}}<1$. The first term in Eq.~\ref{eq:appendix-single-jje-potential-general} is the JJE magnetic energy due to the circulating current $J$ in the JJE loop of inductance $L_{\mathrm{S}}$, which is spontaneously generated in response to the application of the external magnetic flux $\Phi_{\mathrm{ext}}$. This term has previously been omitted in calculations for SNAILs considering the small $\beta_{\mathrm{S}}$ limit~\cite{krantz2013investigation,frattini2017wave,frattini2018optimizing,sivak2019kerr}. The second term in Eq.~\ref{eq:appendix-single-jje-potential-general} is the potential energy of the Josephson junctions. For the particular cases when JJEs are rf SQUIDs, $N=1$, or dc SQUIDs, $N=2$, Eq.~\ref{eq:appendix-single-jje-potential-general} gives the well-known expressions for the potential energy of these systems~\cite{chesca2004squid}.

The phases $\phi_{\mathrm{large}}$ and $\phi_{\mathrm{s}}$ are related via the flux quantization rule within each JJE:

\begin{equation}
\label{eq:appendix-flux-quantization-general}
\begin{split}
(N-1)\phi_{\mathrm{large}}+\phi_{\mathrm{s}}
&=\frac{2\pi}{\Phi_0}\left(\Phi_{\mathrm{ext}}+L_{\mathrm{S}}J\right) \\
&=2\pi\phi_{\mathrm{ext}}+\beta_{\mathrm{S}} j_{\mathrm{n}},
\end{split}
\end{equation}

Using Eq.~\ref{eq:appendix-flux-quantization-general} to eliminate $j_{\mathrm{n}}$ in Eq.~\ref{eq:appendix-single-jje-potential-general}, one can write the two-dimensional JJE potential as:

\begin{equation}
\label{eq:appendix-two-dimensional-potential-general}
\begin{split}
U(\phi_{\mathrm{s}},\phi_{\mathrm{large}})
&=\frac{\alpha}{2\beta_{\mathrm{S}}}E_{\mathrm{J}}\left[(N-1)\phi_{\mathrm{large}}+\phi_{\mathrm{s}}-2\pi\phi_{\mathrm{ext}}\right]^2 \\
&\quad + E_{\mathrm{J}}\left[-\alpha\cos\phi_{\mathrm{s}}-(N-1)\cos\phi_{\mathrm{large}}\right].
\end{split}
\end{equation}

For devices characterized by $\beta_{\mathrm{S}}<0.3$, as a good approximation, we can neglect the $\beta_{\mathrm{S}} j_{\mathrm{n}}$ contribution in Eq.~\ref{eq:appendix-flux-quantization-general}, as well as the first term in Eq.~\ref{eq:appendix-single-jje-potential-general} proportional to $\beta_{\mathrm{S}} j_{\mathrm{n}}^2$. These approximations simplify the problem considerably, because from Eq.~\ref{eq:appendix-flux-quantization-general} we can now eliminate $\phi_{\mathrm{large}}$ and reduce the dimensionality of the problem from a 2D one, $(\phi_{\mathrm{s}},\phi_{\mathrm{large}})$, to a 1D one, $\phi_{\mathrm{s}}$, for each individual JJE:

\begin{subequations}
\label{eq:appendix-large-phase-elimination-general}
\begin{equation}
\label{eq:appendix-large-phase-general}
\phi_{\mathrm{large}}=\frac{2\pi\phi_{\mathrm{ext}}-\phi_{\mathrm{s}}}{N-1}.
\end{equation}
\begin{equation}
\label{eq:appendix-single-jje-potential-reduced}
\begin{split}
U(\phi_{\mathrm{s}})
&=E_{\mathrm{J}}\left[-\alpha\cos\phi_{\mathrm{s}}-(N-1)\cos\left(\frac{2\pi\phi_{\mathrm{ext}}-\phi_{\mathrm{s}}}{N-1}\right)\right].
\end{split}
\end{equation}
\end{subequations}

As in the main text, let us now consider a particular case of an array of $M=M_1+M_2+\ldots+M_P$ unequal JJEs that consists of $P$ different types such that $M_1$ JJEs have an effective area $A_1=R_1A$, $M_2$ JJEs have an effective area $A_2=R_2A$ and so on, and finally $M_P$ JJEs have an effective area $A_P=R_P A$. We denote by $\phi_{\mathrm{s}1},\phi_{\mathrm{s}2},\ldots,\phi_{\mathrm{s}P}$ the superconducting phase differences across the smallest junction in each JJE. Since for arrays of unequal SNAILs the superconducting phase splits unequally among the individual SNAILs, for every applied flux $\Phi_{\mathrm{ext}}$, we have $\phi_{\mathrm{s}1}\neq\phi_{\mathrm{s}2}\neq\ldots\neq\phi_{\mathrm{s}P}$. Using Eq.~\ref{eq:appendix-single-jje-potential-reduced}, the array's $P$-dimensional potential energy can be written as follows, as in Eqs.~\ref{eq:array-potential-sum} and \ref{eq:array-potential-element} of the main text:

\begin{subequations}
\label{eq:appendix-array-potential-general}
\begin{equation}
\label{eq:appendix-array-potential-general-sum}
U_{\mathrm{array}}(\phi_{\mathrm{s}1},\phi_{\mathrm{s}2},\ldots,\phi_{\mathrm{s}P})=\sum_{k=1}^{P}M_k U_k(\phi_{\mathrm{s}k}).
\end{equation}

\begin{equation}
\label{eq:appendix-array-potential-general-element}
\begin{aligned}
U_k(\phi_{\mathrm{s}k})
= E_{\mathrm{J},k}\Bigg[
&-\alpha_k\cos\phi_{\mathrm{s}k} \\
&-(N_k-1)\cos\left(
\frac{R_k 2\pi\phi_{\mathrm{ext}}-\phi_{\mathrm{s}k}}
{N_k-1}
\right)
\Bigg],\\
&\hspace{1.5cm} k=1,2,\ldots,P.
\end{aligned}
\end{equation}

\end{subequations}

Here $R_1\phi_{\mathrm{ext}},R_2\phi_{\mathrm{ext}},\ldots,R_P\phi_{\mathrm{ext}}$ are the normalized externally applied fluxes to individual JJEs with effective areas $A_1,A_2,\ldots,A_P$, respectively. For $N_k=1$, $N_k=2$, and $N_k\geq3$, Eqs.~\ref{eq:appendix-array-potential-general-sum} and \ref{eq:appendix-array-potential-general-element} apply to well-known configurations such as series arrays of unequal rf SQUIDs, unequal dc SQUIDs, or unequal SNAILs, respectively. For the case when JJEs are rf SQUIDs that have just one junction in the loop $(N_k=1)$, only the first term in Eq.~\ref{eq:appendix-array-potential-general-element} should be considered. The multidimensionality feature in the phase space for the potential of arrays of unequal JJEs is in contrast to the case of arrays of identical JJEs that are described by a one-dimensional potential.

\begin{figure}
\centering
\includegraphics[width=\columnwidth]{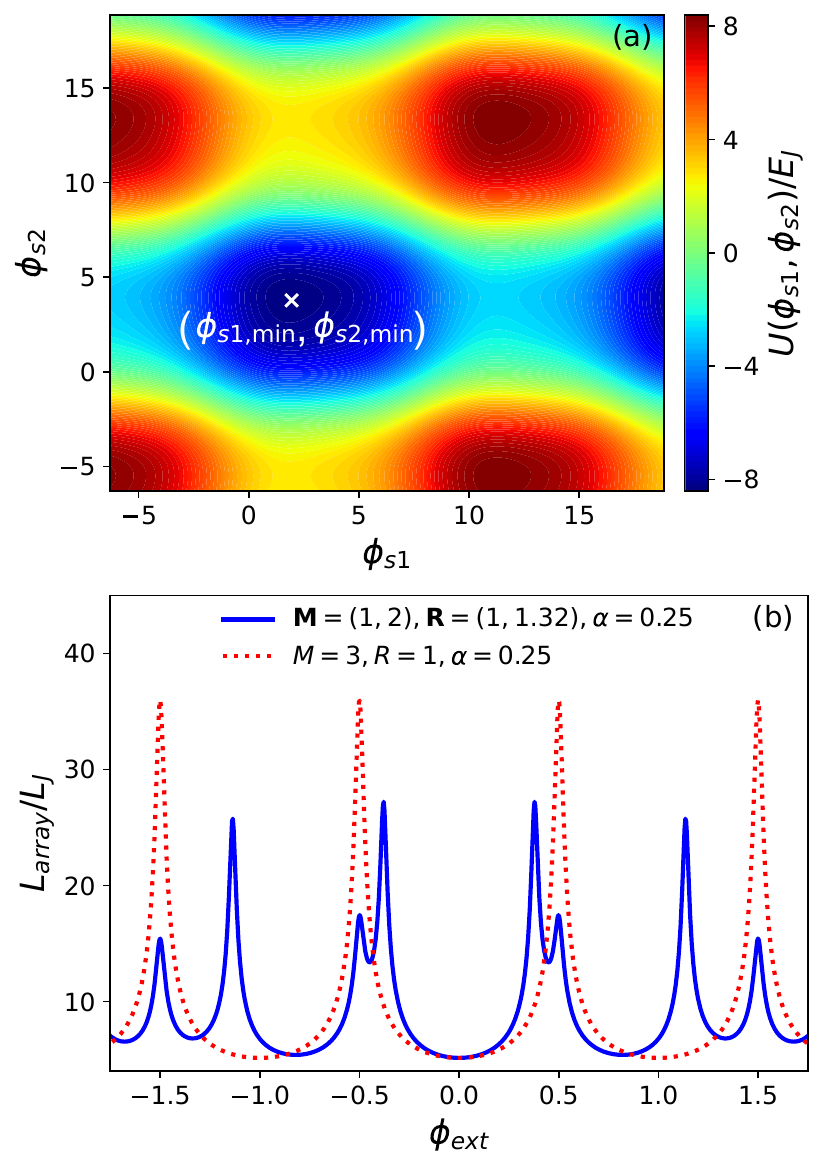}
\caption{(a) Heatmap of the 2D potential $U_{\mathrm{array}}(\phi_{\mathrm{s}1},\phi_{\mathrm{s}2})$ given by Eqs.~\ref{eq:appendix-array-potential-general-sum} and \ref{eq:appendix-array-potential-general-element} for an array of $M=3$ unequal SNAILs, with $\boldsymbol{\mathrm{M}} = (1,2)$, for a particular value of the externally applied flux $\Phi_{\mathrm{ext}}=0.35\Phi_0$; the potential minima are reached at different values for the phases inside the different SNAILs, $\phi_{\mathrm{s}1,\mathrm{min}}\neq\phi_{\mathrm{s}2,\mathrm{min}}$, leading to an unsynchronized operation of various SNAILs in the array. The labeled point indicates one such minimum. (b) The normalized inductance $L_{\mathrm{array}}/L_{\mathrm{J}}$ of an array of unequal SNAILs, with $\boldsymbol{\mathrm{R}} = (1,1.32)$ has a quasi-non-periodic $\Phi_{\mathrm{ext}}$ dependence in sharp contrast to arrays of identical SNAILs, $R=1$, which have a $\Phi_0$ periodicity with $\Phi_{\mathrm{ext}}$.}
\label{fig:a1}
\end{figure}

An example of a heatmap for the particular case of $P=2$ and $N_k=4$, i.e. a two-dimensional potential $U_{\mathrm{array}}(\phi_{\mathrm{s}1},\phi_{\mathrm{s}2})$ for an array of $M=3$ unequal SNAILs, with $\boldsymbol{\mathrm{M}} = (1,2)$, and a particular value of $\Phi_{\mathrm{ext}}=0.35\Phi_0$ is shown in Fig.~\ref{fig:a1}(a). We operate the unequal SNAIL array in one of its equivalent minima. For each $\Phi_{\mathrm{ext}}$ value, one can determine the coordinates of the $U_{\mathrm{array}}(\phi_{\mathrm{s}1},\phi_{\mathrm{s}2})$ minima. As visible from Fig.~\ref{fig:a1}(a), the two coordinates of those minima are different from one another, $\phi_{\mathrm{s}1,\mathrm{min}}\neq\phi_{\mathrm{s}2,\mathrm{min}}$, confirming that the superconducting phase splits unevenly among SNAILs within the array, leading to an unsynchronized operation.

We operate the JJE array in one of its equivalent minima. For each $\Phi_{\mathrm{ext}}$ value, one can determine the coordinates of the $U_{\mathrm{array}}(\phi_{\mathrm{s}1},\phi_{\mathrm{s}2},\ldots,\phi_{\mathrm{s}P})$ minima, $(\phi_{\mathrm{s}1,\mathrm{min}},\phi_{\mathrm{s}2,\mathrm{min}},\ldots,\phi_{\mathrm{s}P,\mathrm{min}})$, by solving the following $P$ equations:

\begin{equation}
\label{eq:appendix-array-minimum-condition}
\frac{\partial U_{\mathrm{array}}}{\partial\phi_{\mathrm{s}k}}=M_k\frac{\mathrm{d}U_k}{\mathrm{d}\phi_{\mathrm{s}k}}=\frac{\Phi_0}{2\pi}M_k I_k=0,\qquad k=1,2,\ldots,P.
\end{equation}

Equation~\ref{eq:appendix-array-minimum-condition} is precisely the current $I_k=0$, with $k=1,2,\ldots,P$, conservation condition at each node between different individual JJEs. Generally speaking, the values of all $P$ coordinates of those minima are different from one another, $\phi_{\mathrm{s}1,\mathrm{min}}\neq\phi_{\mathrm{s}2,\mathrm{min}}\neq\ldots\neq\phi_{\mathrm{s}P,\mathrm{min}}$. As an example, for an array of $M=3$ unequal SNAILs, with $\boldsymbol{\mathrm{M}} = (1,2)$, and $P=2$, the system is characterized by a 2D potential shown in Fig.~\ref{fig:a1}(a), whose minima $U_{\mathrm{array}}(\phi_{\mathrm{s}1,\mathrm{min}},\phi_{\mathrm{s}2,\mathrm{min}})$ are found by solving two equations of the type Eq.~\ref{eq:appendix-array-minimum-condition}.

The JPA participation ratio $p$ is defined as:

\begin{equation}
\label{eq:appendix-participation-ratio}
p(\phi_{\mathrm{ext}})=\frac{L_{\mathrm{array}}(\phi_{\mathrm{ext}})}{L+L_{\mathrm{array}}(\phi_{\mathrm{ext}})}.
\end{equation}

Here $L_{\mathrm{array}}$ is given by Eqs.~\ref{eq:array-inductance-total} and \ref{eq:josephson-inductance} from the main text. Unlike the case of identical JJE arrays, whose $L_{\mathrm{array}}$ has a $\Phi_0$-periodic dependence with $\phi_{\mathrm{ext}}$, for arrays made of unequal JJEs, both $L_{\mathrm{array}}$ and the participation ratio $p$ have a quasi-non-periodic dependence. When JJEs are SNAILs, with $N=4$, $L_{\mathrm{array}}(\phi_{\mathrm{ext}})$ calculated with Eqs.~\ref{eq:array-inductance-total} and \ref{eq:josephson-inductance} is shown in Fig.~\ref{fig:a1}(b) for two particular cases: an array consisting of $M=3$ nominally identical SNAILs and an array made of unequal SNAILs with $P=2$, and $\boldsymbol{\mathrm{M}} = (1,2)$.

To implement a JPA, an array of unequal JJEs is embedded into a $\lambda/2$ superconducting transmission line resonator. We work in the frame of the single-mode lumped-element model consisting of a series combination of capacitance $C_{\rm JPA}$, inductance $L$ of the $\lambda/2$ transmission line, and the inductance $L_{\mathrm{array}}$ of the SNAIL array. The total potential energy for the JPA is given by:

\begin{equation}
\label{eq:appendix-jpa-potential-general}
\begin{split}
U_{\mathrm{JPA}}(\phi_{\mathrm{s}1},\phi_{\mathrm{s}2},\ldots,\phi_{\mathrm{s}P},\phi)
&=\frac{1}{2\beta}E_{\mathrm{J}}\left(\phi-\sum_{k=1}^{P}M_k\phi_{\mathrm{s}k}\right)^2 \\
&\quad + U_{\mathrm{array}}(\phi_{\mathrm{s}1},\phi_{\mathrm{s}2},\ldots,\phi_{\mathrm{s}P}).
\end{split}
\end{equation}

Here $\phi$ is the mode canonical phase coordinate, $\beta=2\pi L I_{\mathrm{c}}/\Phi_0$, and $U_{\mathrm{array}}$ is given by Eqs.~\ref{eq:appendix-array-potential-general-sum} and \ref{eq:appendix-array-potential-general-element}. Since the coordinates $\phi_{\mathrm{s}1},\phi_{\mathrm{s}2},\ldots,\phi_{\mathrm{s}P}$ are not independent variables, we can eliminate them by minimizing the potential $U_{\mathrm{JPA}}(\phi_{\mathrm{s}1},\phi_{\mathrm{s}2},\ldots,\phi_{\mathrm{s}P},\phi)$ with respect to $\phi_{\mathrm{s}1},\phi_{\mathrm{s}2},\ldots,\phi_{\mathrm{s}P}$, as in the main text. In doing so the problem becomes 1D: $U_{\mathrm{JPA}}=U_{\mathrm{JPA}}(\phi)$.

In the following, for simplicity, we assume $P=2$, $\mathbf{M} = (M_1, M_2)$ and $\mathbf{R} = (1, R_2)$ so that the JPA potential is given by particular cases of Eqs.~\ref{eq:appendix-jpa-potential-general}, \ref{eq:appendix-array-potential-general-sum}, and \ref{eq:appendix-array-potential-general-element}. The calculations that follow can easily be generalized for the case when $P>2$:

\begin{subequations}
\label{eq:appendix-p2-jpa-potential}
\begin{equation}
\label{eq:appendix-p2-jpa-potential-total}
\begin{split}
U_{\mathrm{JPA}}(\phi_{\mathrm{s}1},\phi_{\mathrm{s}2},\phi)
&=\frac{1}{2\beta}E_{\mathrm{J}}\left(\phi-M_1\phi_{\mathrm{s}1}-M_2\phi_{\mathrm{s}2}\right)^2 \\
&\quad + U_{\mathrm{array}}(\phi_{\mathrm{s}1},\phi_{\mathrm{s}2}).
\end{split}
\end{equation}
\begin{equation}
\label{eq:appendix-p2-array-potential}
U_{\mathrm{array}}(\phi_{\mathrm{s}1},\phi_{\mathrm{s}2})=M_1U_1(\phi_{\mathrm{s}1})+M_2U_2(\phi_{\mathrm{s}2}).
\end{equation}
\begin{equation}
\label{eq:appendix-p2-u1}
U_1(\phi_{\mathrm{s}1})=E_{\mathrm{J}}\left[-\alpha\cos\phi_{\mathrm{s}1}-3\cos\left(\frac{2\pi\phi_{\mathrm{ext}}-\phi_{\mathrm{s}1}}{3}\right)\right].
\end{equation}
\begin{equation}
\label{eq:appendix-p2-u2}
U_2(\phi_{\mathrm{s}2})=E_{\mathrm{J}}\left[-\alpha\cos\phi_{\mathrm{s}2}-3\cos\left(\frac{2\pi R_2\phi_{\mathrm{ext}}-\phi_{\mathrm{s}2}}{3}\right)\right].
\end{equation}
\end{subequations}

As mentioned, coordinates $\phi_{\mathrm{s}1}$ and $\phi_{\mathrm{s}2}$ are not independent variables, so we have to eliminate them by minimizing the potential $U(\phi_{\mathrm{s}1},\phi_{\mathrm{s}2},\phi)$ with respect to both $\phi_{\mathrm{s}1}$ and $\phi_{\mathrm{s}2}$. This is equivalent to imposing two current conservation conditions: one at the node between one type of SNAIL and the linear inductor and the second one at the virtual node between the second type of SNAIL and the linear inductor:

\begin{subequations}
\label{eq:appendix-current-conservation}
\begin{equation}
\label{eq:appendix-current-conservation-1}
\begin{split}
0&=\frac{1}{M_1}\frac{\partial U_{\mathrm{JPA}}}{\partial\phi_{\mathrm{s}1}} \\
&=-E_{\mathrm{L}}(\phi-M_1\phi_{\mathrm{s}1}-M_2\phi_{\mathrm{s}2})+\frac{\mathrm{d}U_1}{\mathrm{d}\phi_{\mathrm{s}1}}.
\end{split}
\end{equation}
\begin{equation}
\label{eq:appendix-current-conservation-2}
\begin{split}
0&=\frac{1}{M_2}\frac{\partial U_{\mathrm{JPA}}}{\partial\phi_{\mathrm{s}2}} \\
&=-E_{\mathrm{L}}(\phi-M_1\phi_{\mathrm{s}1}-M_2\phi_{\mathrm{s}2})+\frac{\mathrm{d}U_2}{\mathrm{d}\phi_{\mathrm{s}2}}.
\end{split}
\end{equation}
\end{subequations}

Here $E_{\mathrm{L}} = \frac{E_{\mathrm{J}}}{\beta} = \frac{1}{L}\left(\frac{\Phi_0}{2\pi}\right)^2
$. Solving the system of two coupled simultaneous equations, Eqs.~\ref{eq:appendix-current-conservation-1} and \ref{eq:appendix-current-conservation-2}, we find $\phi_{\mathrm{s}1}(\phi)$ and $\phi_{\mathrm{s}2}(\phi)$ so that the JPA total potential energy becomes a one-variable function $U_{\mathrm{JPA}}=U_{\mathrm{JPA}}(\phi)$.

We operate the JPA in one of its equivalent equilibrium minima $\phi_{\mathrm{min}}$ given by:

\begin{subequations}
\label{eq:appendix-taylor-expansion-general}
\begin{equation}
\label{eq:appendix-equilibrium-condition-general}
c_1=\frac{1}{E_{\mathrm{J}}}\left.\frac{\mathrm{d}U_{\mathrm{JPA}}}{\mathrm{d}\phi}\right|_{\phi_{min}}=0.
\end{equation}
\begin{equation}
\label{eq:appendix-taylor-potential-general}
\begin{split}
U_{\mathrm{JPA}}(\phi)
&=E_{\mathrm{J}}\left[\frac{c_2}{2!}(\phi-\phi_{\mathrm{min}})^2+\frac{c_3}{3!}(\phi-\phi_{\mathrm{min}})^3+\ldots\right].
\end{split}
\end{equation}
\begin{equation}
\label{eq:appendix-taylor-coefficients-general}
c_n=\frac{1}{E_{\mathrm{J}}}\left.\frac{\mathrm{d}^n U_{\mathrm{JPA}}}{\mathrm{d}\phi^n}\right|_{\phi_{\mathrm{min}}},\qquad n=2,3,4,\ldots.
\end{equation}
\end{subequations}

Using Eqs.~\ref{eq:appendix-p2-jpa-potential-total} and \ref{eq:appendix-p2-array-potential} in Eq.~\ref{eq:appendix-taylor-coefficients-general}, we find for the first three non-zero coefficients the following expressions (evaluated for $P=2)$:

\begin{subequations}
\label{eq:appendix-taylor-coefficients-explicit}
\begin{equation}
\label{eq:appendix-c2-coefficient}
c_2(\phi_{\mathrm{ext}})=\frac{1}{\beta}\left[1-M_1\left.\frac{\mathrm{d}\phi_{\mathrm{s}1}}{\mathrm{d}\phi}\right|_{\phi_{\mathrm{min}}}-M_2\left.\frac{\mathrm{d}\phi_{\mathrm{s}2}}{\mathrm{d}\phi}\right|_{\phi_{\mathrm{min}}}\right]
\end{equation}
\begin{equation}
\label{eq:appendix-c3-coefficient}
c_3(\phi_{\mathrm{ext}})=\frac{1}{\beta}\left[-M_1\left.\frac{\mathrm{d}^2\phi_{\mathrm{s}1}}{\mathrm{d}\phi^2}\right|_{\phi_{\mathrm{min}}}-M_2\left.\frac{\mathrm{d}^2\phi_{\mathrm{s}2}}{\mathrm{d}\phi^2}\right|_{\phi_{\mathrm{min}}}\right]
\end{equation}
\begin{equation}
\label{eq:appendix-c4-coefficient}
c_4(\phi_{\mathrm{ext}})=\frac{1}{\beta}\left[-M_1\left.\frac{\mathrm{d}^3\phi_{\mathrm{s}1}}{\mathrm{d}\phi^3}\right|_{\phi_{\mathrm{min}}}-M_2\left.\frac{\mathrm{d}^3\phi_{\mathrm{s}2}}{\mathrm{d}\phi^3}\right|_{\phi_{\mathrm{min}}}\right]
\end{equation}
\end{subequations}

The derivatives of the implicit functions $\phi_{\mathrm{s}1}(\phi)$ and $\phi_{\mathrm{s}2}(\phi)$ in Eqs.~\ref{eq:appendix-c2-coefficient}--\ref{eq:appendix-c4-coefficient} can be found by differentiating Eqs.~\ref{eq:appendix-current-conservation-1} and \ref{eq:appendix-current-conservation-2}. We can generalise Eqs.~\ref{eq:appendix-taylor-coefficients-explicit} for $P>2$ as:

\begin{subequations}
\label{eq:appendix-c-coefficients_general}

\begin{equation}
\label{eq:appendix-c2-coefficient_general}
c_2(\phi_{\mathrm{ext}})
= \frac{1}{\beta}
\left[
1-\sum_{k=1}^{P}M_k
\left.
\frac{\mathrm{d}\phi_{\mathrm{s}k}}{\mathrm{d}\phi}
\right|_{\phi_{\mathrm{min}}}
\right]
\end{equation}

\begin{equation}
\label{eq:appendix-c3-coefficient_general}
c_3(\phi_{\mathrm{ext}})
= \frac{1}{\beta}
\left[
-\sum_{k=1}^{P}M_k
\left.
\frac{\mathrm{d}^2\phi_{\mathrm{s}k}}{\mathrm{d}\phi^2}
\right|_{\phi_{\mathrm{min}}}
\right]
\end{equation}

\begin{equation}
\label{eq:appendix-c4-coefficient_general}
c_4(\phi_{\mathrm{ext}})
= \frac{1}{\beta}
\left[
-\sum_{k=1}^{P}M_k
\left.
\frac{\mathrm{d}^3\phi_{\mathrm{s}k}}{\mathrm{d}\phi^3}
\right|_{\phi_{\mathrm{min}}}
\right]
\end{equation}

\end{subequations}

For example, to calculate $|g_3\alpha_{\mathrm{p},\mathrm{eff}}|^2$ given by Eq.~\ref{eq:effective-pump} and plotted in the main text, we use the explicit expressions of the $c_2$ and $c_3$ coefficients given by Eqs.~\ref{eq:appendix-c2-coefficient} and \ref{eq:appendix-c3-coefficient} in Fig.~\ref{fig:2}(b) and Eqs.~\ref{eq:appendix-c2-coefficient_general} and \ref{eq:appendix-c3-coefficient_general} setting $P=3$ for Figs.~\ref{fig:5}(b,d). Then the derivatives in  Eqs.~\ref{eq:appendix-c2-coefficient} and \ref{eq:appendix-c3-coefficient} can be calculated as: 

\begin{subequations}
\label{eq:appendix-implicit-derivatives}
\begin{equation}
\label{eq:appendix-dphi-s1-dphi}
\frac{\mathrm{d}\phi_{\mathrm{s}1}}{\mathrm{d}\phi}=\left[M_1+M_2\frac{\mathrm{d}^2U_1/\mathrm{d}\phi_{\mathrm{s}1}^2}{\mathrm{d}^2U_2/\mathrm{d}\phi_{\mathrm{s}2}^2}+\frac{1}{E_{\mathrm{L}}}\frac{\mathrm{d}^2U_1}{\mathrm{d}\phi_{\mathrm{s}1}^2}\right]^{-1}.
\end{equation}
\begin{equation}
\label{eq:appendix-dphi-s2-dphi}
\frac{\mathrm{d}\phi_{\mathrm{s}2}}{\mathrm{d}\phi}=\left[M_2+M_1\frac{\mathrm{d}^2U_2/\mathrm{d}\phi_{\mathrm{s}2}^2}{\mathrm{d}^2U_1/\mathrm{d}\phi_{\mathrm{s}1}^2}+\frac{1}{E_{\mathrm{L}}}\frac{\mathrm{d}^2U_2}{\mathrm{d}\phi_{\mathrm{s}2}^2}\right]^{-1}.
\end{equation}
\begin{equation}
\label{eq:appendix-d2phi-s1-dphi2}
\begin{split}
\frac{\mathrm{d}^2\phi_{\mathrm{s}1}}{\mathrm{d}\phi^2}
&=-M_2\left(\frac{\mathrm{d}\phi_{\mathrm{s}1}}{\mathrm{d}\phi}\right)^2\frac{\mathrm{d}}{\mathrm{d}\phi}\left(\frac{\mathrm{d}^2U_1/\mathrm{d}\phi_{\mathrm{s}1}^2}{\mathrm{d}^2U_2/\mathrm{d}\phi_{\mathrm{s}2}^2}\right) \\
&\quad -\frac{1}{E_{\mathrm{L}}}\frac{\mathrm{d}^3U_1}{\mathrm{d}\phi_{\mathrm{s}1}^3}\left(\frac{\mathrm{d}\phi_{\mathrm{s}1}}{\mathrm{d}\phi}\right)^3.
\end{split}
\end{equation}
\begin{equation}
\label{eq:appendix-d2phi-s2-dphi2}
\begin{split}
\frac{\mathrm{d}^2\phi_{\mathrm{s}2}}{\mathrm{d}\phi^2}
&=-M_1\left(\frac{\mathrm{d}\phi_{\mathrm{s}2}}{\mathrm{d}\phi}\right)^2\frac{\mathrm{d}}{\mathrm{d}\phi}\left(\frac{\mathrm{d}^2U_2/\mathrm{d}\phi_{\mathrm{s}2}^2}{\mathrm{d}^2U_1/\mathrm{d}\phi_{\mathrm{s}1}^2}\right) \\
&\quad -\frac{1}{E_{\mathrm{L}}}\frac{\mathrm{d}^3U_2}{\mathrm{d}\phi_{\mathrm{s}2}^3}\left(\frac{\mathrm{d}\phi_{\mathrm{s}2}}{\mathrm{d}\phi}\right)^3.
\end{split}
\end{equation}
\end{subequations}

Finally, the explicit expressions of the first term on the right-hand side of Eqs.~\ref{eq:appendix-d2phi-s1-dphi2} and \ref{eq:appendix-d2phi-s2-dphi2} are given by:

\begin{subequations}
\label{eq:appendix-derivative-ratios}
\begin{equation}
\label{eq:appendix-derivative-ratio-1-over-2}
\begin{split}
&\frac{\mathrm{d}}{\mathrm{d}\phi}\left(\frac{\mathrm{d}^2U_1/\mathrm{d}\phi_{\mathrm{s}1}^2}{\mathrm{d}^2U_2/\mathrm{d}\phi_{\mathrm{s}2}^2}\right)
=\frac{\dfrac{\mathrm{d}^3U_1}{\mathrm{d}\phi_{\mathrm{s}1}^3}\dfrac{\mathrm{d}^2U_2}{\mathrm{d}\phi_{\mathrm{s}2}^2}\dfrac{\mathrm{d}\phi_{\mathrm{s}1}}{\mathrm{d}\phi}-\dfrac{\mathrm{d}^3U_2}{\mathrm{d}\phi_{\mathrm{s}2}^3}\dfrac{\mathrm{d}^2U_1}{\mathrm{d}\phi_{\mathrm{s}1}^2}\dfrac{\mathrm{d}\phi_{\mathrm{s}2}}{\mathrm{d}\phi}}{\left(\mathrm{d}^2U_2/\mathrm{d}\phi_{\mathrm{s}2}^2\right)^2}.
\end{split}
\end{equation}
\begin{equation}
\label{eq:appendix-derivative-ratio-2-over-1}
\begin{split}
&\frac{\mathrm{d}}{\mathrm{d}\phi}\left(\frac{\mathrm{d}^2U_2/\mathrm{d}\phi_{\mathrm{s}2}^2}{\mathrm{d}^2U_1/\mathrm{d}\phi_{\mathrm{s}1}^2}\right)
=\frac{\dfrac{\mathrm{d}^3U_2}{\mathrm{d}\phi_{\mathrm{s}2}^3}\dfrac{\mathrm{d}^2U_1}{\mathrm{d}\phi_{\mathrm{s}1}^2}\dfrac{\mathrm{d}\phi_{\mathrm{s}2}}{\mathrm{d}\phi}-\dfrac{\mathrm{d}^3U_1}{\mathrm{d}\phi_{\mathrm{s}1}^3}\dfrac{\mathrm{d}^2U_2}{\mathrm{d}\phi_{\mathrm{s}2}^2}\dfrac{\mathrm{d}\phi_{\mathrm{s}1}}{\mathrm{d}\phi}}{\left(\mathrm{d}^2U_1/\mathrm{d}\phi_{\mathrm{s}1}^2\right)^2}.
\end{split}
\end{equation}
\end{subequations}

The derivatives required to calculate $c_4$ given by Eq.~\ref{eq:appendix-c4-coefficient} can straightforwardly be calculated in a similar fashion.

To optimize operation of JPAs based on arrays of unequal SNAILs, we need to determine the expressions of the $g_3$ and $g_4$ coefficients, responsible for gain in the three-wave-mixing and four-wave-mixing regimes, respectively, as given by Eq.~\ref{eq:nonlinearity-coefficients}:

\begin{subequations}
\label{eq:appendix-g-coefficients}
\begin{equation}
\label{eq:appendix-g3-coefficient}
g_3=\frac{1}{3!}\frac{c_3}{c_2}\left(\frac{\omega_{\mathrm{a}}E_{\mathrm{C}}}{\hbar}\right)^{1/2}.
\end{equation}
\begin{equation}
\label{eq:appendix-g4-coefficient}
g_4=\frac{1}{4!}\frac{c_4}{c_2}\left(\frac{2E_{\mathrm{C}}}{\hbar}\right).
\end{equation}
\end{subequations}

Taking into account the inhomogeneous normalised rf pump across the array, the terms $|g_3\alpha_{\mathrm{p},\mathrm{eff}}|^2$ and $|g_4\alpha_{\mathrm{p},\mathrm{eff}}^2|^2$ are the most relevant quantities defining gain $G$ and bandwidth in arrays of unequal-SNAIL-based JPAs:

\begin{subequations}
\label{eq:appendix-effective-pump}
\begin{equation}
\label{eq:appendix-g3-effective-pump}
\begin{split}
\left|g_3 \alpha_{\mathrm{p},\mathrm{eff}}\right|^2
&=\left|g_3\alpha_{\mathrm{p}}\frac{L_{\mathrm{array}}}{L_{\mathrm{J}}}\right|^2 \\
&=\left(\frac{1}{3!}\alpha_{\mathrm{p}}\frac{1}{L_{\mathrm{J}}}\right)^2\frac{E_{\mathrm{C}}}{\hbar}\left|\frac{c_3}{c_2}\sqrt{\omega_{\mathrm{a}}}L_{\mathrm{array}}\right|^2.
\end{split}
\end{equation}
\begin{equation}
\label{eq:appendix-g4-effective-pump}
\begin{split}
\left|g_4 \alpha_{\mathrm{p},\mathrm{eff}}^2\right|^2
&=\left|g_4\alpha_{\mathrm{p}}^2\left(\frac{L_{\mathrm{array}}}{L_{\mathrm{J}}}\right)^2\right|^2 \\
&=\left(\frac{1}{4!}\alpha_{\mathrm{p}}^2\frac{1}{L_{\mathrm{J}}^2}\right)^2\left(\frac{2E_{\mathrm{C}}}{\hbar}\right)^2\left|\frac{c_4}{c_2}L_{\mathrm{array}}^2\right|^2.
\end{split}
\end{equation}
\end{subequations}

\begin{figure}
\centering
\includegraphics[width=\columnwidth]{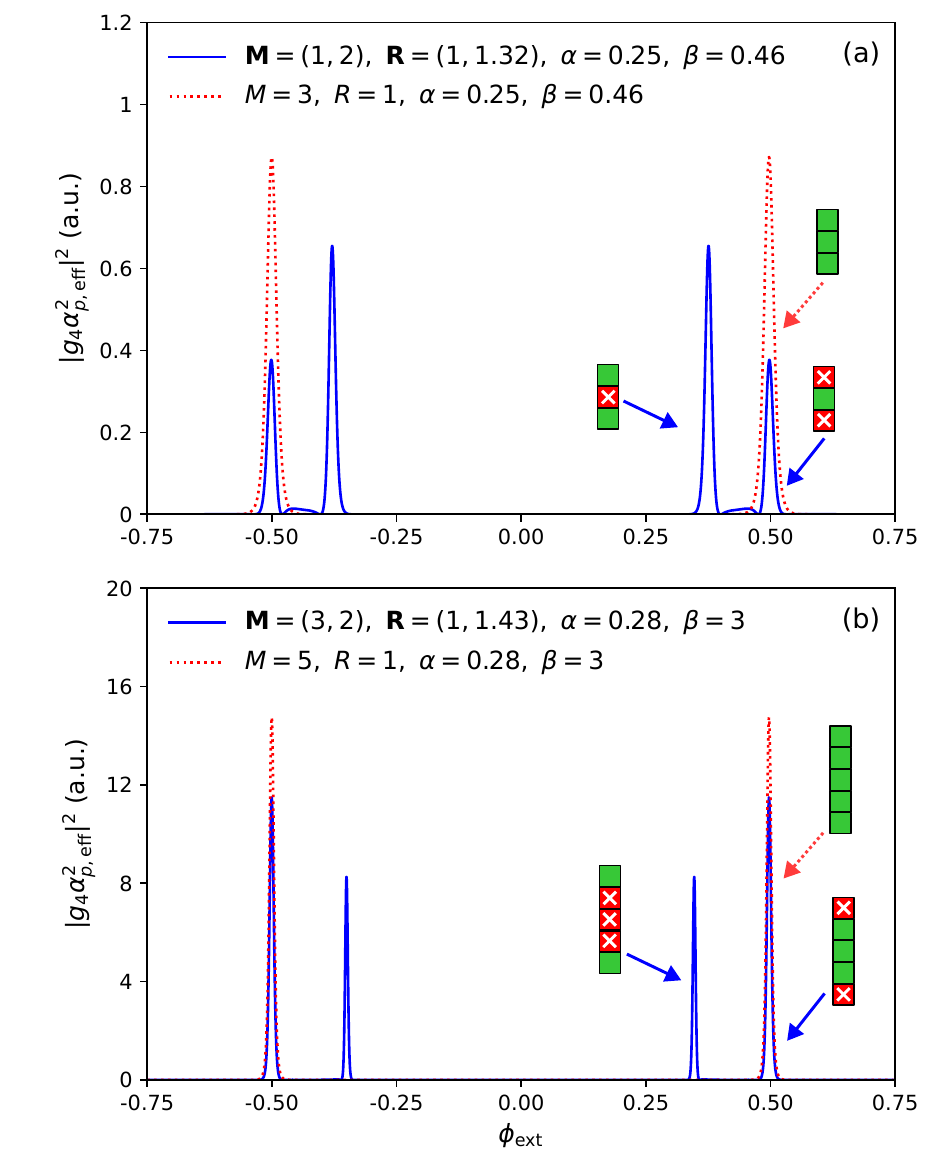}
\caption{Calculated $|g_4\alpha_{\mathrm{p},\mathrm{eff}}^2|^2(\phi_{\mathrm{ext}})$ using Eq.~\ref{eq:appendix-g4-effective-pump} for (a) an array of unequal SNAILs with $\boldsymbol{\mathrm{M}} = (1,2)$ and an array of $M=3$ identical SNAILs, and (b) an array of unequal SNAILs with $\boldsymbol{\mathrm{M}} = (3,2)$ and an array of $M=5$ identical SNAILs. We reproduce the green/red with superimposed X squares as shown in Fig.~\ref{fig:2} and Fig.~\ref{fig:5} to indicate the JJEs which are active amplifiers or lumped inductors respectively. We illustrate the physics here using a simple case of two values of flux focusing of SNAIL, whereas the experimental demonstration in the main text has three values.}
\label{fig:a2}
\end{figure}

All terms inside the modulus functions in Eqs.~\ref{eq:appendix-g3-effective-pump} and \ref{eq:appendix-g4-effective-pump} have an explicit dependence on $\phi_{\mathrm{ext}}$. Using the explicit flux dependencies of $c_2$, $c_3$, $c_4$, $\omega_{\mathrm{a}}$, and $L_{\mathrm{array}}$ determined above, we can now use Eqs.~\ref{eq:appendix-g3-effective-pump} and \ref{eq:appendix-g4-effective-pump} to plot the flux dependencies of $|g_3\alpha_{\mathrm{p},\mathrm{eff}}|^2$ in Fig.~\ref{fig:2}(b), Fig.~\ref{fig:5}(b, d) and $|g_4\alpha_{\mathrm{p},\mathrm{eff}}^2|^2$ (Fig.~\ref{fig:a2}) and choose the optimal flux bias points for both operating regimes, three-wave mixing and four-wave mixing.

\section{Arrays of unequal JJEs, the concept}

Generally speaking, an array of unequal JJEs consists of at least two different JJEs. There are many ways in practice to implement an array of unequal JJEs. Here two possibilities are explored when the JJEs are different in the sense that:

\begin{enumerate}
\item They are purposely flux-biased differently, i.e. the flux coupled to each SNAIL may differ because $A_i\neq A_j$ for some or all $i\neq j$.
\item Although JJEs are flux-biased identically, the phase drop across each JJE may differ, $\phi_i\neq\phi_j$ for some or all $i\neq j$. This possibility can be implemented in practice in many ways: either fabricating JJEs having flux quantization loops with variable inner perimeter $P_i$, fabricating variable $\alpha_i$ JJEs, fabricating variable Josephson critical current $I_i$ JJEs, having JJEs with varied $N_k$ combining rf, dc SQUIDs, SNAILs and other types of JJE or combinations of these different approaches.
\item The two approaches above can also be combined.
\end{enumerate}

Moreover, there are multiple ways in practice to implement an array of $M$ unequal JJEs. For example, some but not all JJEs may be different, or all JJEs may be different.

For arrays made of nominally identical JJEs, the superconducting phase splits evenly among all JJEs, $\phi_i=\phi_j$ for all $i\neq j$ with $1\leq i,j\leq M$, where $\phi_i$ and $\phi_j$ are the phase drops across JJE $i$ and JJE $j$, respectively, while all JJEs have identical effective areas, $A_i=A_j$. Consequently, the arrays are described by a one-dimensional Hamiltonian that has a flux-quantum periodicity, $\Phi_0$, with $\Phi_{\mathrm{ext}}$.

In contrast, for arrays made of unequal JJEs the total phase splits unevenly among them, so that $\phi_i\neq\phi_j$ with $i\neq j$ each time we compare two unequal JJEs. This is the case, in particular, when JJEs have different effective areas $A_i\neq A_j$, with $i\neq j$, so that the external flux coupled individually to various unequal JJEs differs: $\Phi_{\mathrm{ext},i}\neq\Phi_{\mathrm{ext},j}$. As a result, the potential $U$ describing the system becomes multidimensional and quasi-non-periodic with $\Phi_{\mathrm{ext}}$. Here we define $\Phi_{\mathrm{ext}}$ as the applied flux to JJEs with the smallest effective area. The term quasi-non-periodic should be understood in the sense that the system may show:

\begin{enumerate}
\item No periodicity at all with $\Phi_{\mathrm{ext}}$, in the case where the ratio $A_i/A_j$ is not a rational number $m/n$, with $m$ and $n$ being positive integers.
\item A larger-than-$\Phi_0$ periodicity with $\Phi_{\mathrm{ext}}$, if $A_i/A_j$ is a rational number $m/n\geq1$.
\end{enumerate}

Framed in terms of flux-focusing ratios $R$, when all $R_k$ are rational and expressed relative to a reference $R_1=1$, a common period of the individual flux biases is determined by the least common multiple of their reduced denominators, subject to possible additional symmetries or branch-selection effects. For the two examples implemented experimentally in the main text, $\mathbf{R}=(1, 1.32, 1.40)$ and $\mathbf{R}=(1, 1.43, 1.60)$ the periodicity (assuming these values to be exact) is 25 $\Phi_0$ and 100 $\Phi_0$ respectively.

\begin{acknowledgments}
We thank the members of the OQC Nanofabrication and Materials teams for discussions relating to the manufacture of the amplifiers, and the OQC Laboratory team for their help with the cryostats. We thank Apostolous Koukounaris for manufacture of the packages. We thank Brian Vlastakis and Thomas Sweetnam for useful discussions throughout this project. We thank Gioele Consani, Brian Vlastakis, and Peter Leek for providing critical feedback on this manuscript. We thank the Royal Holloway University of London SuperFab Facility for their support.
\end{acknowledgments}

\bibliography{bibliography}

\begin{thebibliography}{40}%
\makeatletter
\providecommand \@ifxundefined [1]{%
 \@ifx{#1\undefined}
}%
\providecommand \@ifnum [1]{%
 \ifnum #1\expandafter \@firstoftwo
 \else \expandafter \@secondoftwo
 \fi
}%
\providecommand \@ifx [1]{%
 \ifx #1\expandafter \@firstoftwo
 \else \expandafter \@secondoftwo
 \fi
}%
\providecommand \natexlab [1]{#1}%
\providecommand \enquote  [1]{``#1''}%
\providecommand \bibnamefont  [1]{#1}%
\providecommand \bibfnamefont [1]{#1}%
\providecommand \citenamefont [1]{#1}%
\providecommand \href@noop [0]{\@secondoftwo}%
\providecommand \href [0]{\begingroup \@sanitize@url \@href}%
\providecommand \@href[1]{\@@startlink{#1}\@@href}%
\providecommand \@@href[1]{\endgroup#1\@@endlink}%
\providecommand \@sanitize@url [0]{\catcode `\\12\catcode `\$12\catcode `\&12\catcode `\#12\catcode `\^12\catcode `\_12\catcode `\%12\relax}%
\providecommand \@@startlink[1]{}%
\providecommand \@@endlink[0]{}%
\providecommand \url  [0]{\begingroup\@sanitize@url \@url }%
\providecommand \@url [1]{\endgroup\@href {#1}{\urlprefix }}%
\providecommand \urlprefix  [0]{URL }%
\providecommand \Eprint [0]{\href }%
\providecommand \doibase [0]{https://doi.org/}%
\providecommand \selectlanguage [0]{\@gobble}%
\providecommand \bibinfo  [0]{\@secondoftwo}%
\providecommand \bibfield  [0]{\@secondoftwo}%
\providecommand \translation [1]{[#1]}%
\providecommand \BibitemOpen [0]{}%
\providecommand \bibitemStop [0]{}%
\providecommand \bibitemNoStop [0]{.\EOS\space}%
\providecommand \EOS [0]{\spacefactor3000\relax}%
\providecommand \BibitemShut  [1]{\csname bibitem#1\endcsname}%
\let\auto@bib@innerbib\@empty
\bibitem [{\citenamefont {Yamamoto}\ \emph {et~al.}(2008)\citenamefont {Yamamoto}, \citenamefont {Inomata}, \citenamefont {Watanabe}, \citenamefont {Matsuba}, \citenamefont {Miyazaki}, \citenamefont {Oliver}, \citenamefont {Nakamura},\ and\ \citenamefont {Tsai}}]{yamamoto2008flux}%
  \BibitemOpen
  \bibfield  {author} {\bibinfo {author} {\bibfnamefont {T.}~\bibnamefont {Yamamoto}}, \bibinfo {author} {\bibfnamefont {K.}~\bibnamefont {Inomata}}, \bibinfo {author} {\bibfnamefont {M.}~\bibnamefont {Watanabe}}, \bibinfo {author} {\bibfnamefont {K.}~\bibnamefont {Matsuba}}, \bibinfo {author} {\bibfnamefont {T.}~\bibnamefont {Miyazaki}}, \bibinfo {author} {\bibfnamefont {W.~D.}\ \bibnamefont {Oliver}}, \bibinfo {author} {\bibfnamefont {Y.}~\bibnamefont {Nakamura}},\ and\ \bibinfo {author} {\bibfnamefont {J.}~\bibnamefont {Tsai}},\ }\bibfield  {title} {\bibinfo {title} {Flux-driven {Josephson} parametric amplifier},\ }\href@noop {} {\bibfield  {journal} {\bibinfo  {journal} {Applied Physics Letters}\ }\textbf {\bibinfo {volume} {93}},\ \bibinfo {pages} {042510} (\bibinfo {year} {2008})}\BibitemShut {NoStop}%
\bibitem [{\citenamefont {Krantz}\ \emph {et~al.}(2016)\citenamefont {Krantz}, \citenamefont {Bengtsson}, \citenamefont {Simoen}, \citenamefont {Gustavsson}, \citenamefont {Shumeiko}, \citenamefont {Oliver}, \citenamefont {Wilson}, \citenamefont {Delsing},\ and\ \citenamefont {Bylander}}]{krantz2016single}%
  \BibitemOpen
  \bibfield  {author} {\bibinfo {author} {\bibfnamefont {P.}~\bibnamefont {Krantz}}, \bibinfo {author} {\bibfnamefont {A.}~\bibnamefont {Bengtsson}}, \bibinfo {author} {\bibfnamefont {M.}~\bibnamefont {Simoen}}, \bibinfo {author} {\bibfnamefont {S.}~\bibnamefont {Gustavsson}}, \bibinfo {author} {\bibfnamefont {V.}~\bibnamefont {Shumeiko}}, \bibinfo {author} {\bibfnamefont {W.}~\bibnamefont {Oliver}}, \bibinfo {author} {\bibfnamefont {C.}~\bibnamefont {Wilson}}, \bibinfo {author} {\bibfnamefont {P.}~\bibnamefont {Delsing}},\ and\ \bibinfo {author} {\bibfnamefont {J.}~\bibnamefont {Bylander}},\ }\bibfield  {title} {\bibinfo {title} {Single-shot read-out of a superconducting qubit using a {Josephson} parametric oscillator},\ }\href@noop {} {\bibfield  {journal} {\bibinfo  {journal} {Nature communications}\ }\textbf {\bibinfo {volume} {7}},\ \bibinfo {pages} {11417} (\bibinfo {year} {2016})}\BibitemShut {NoStop}%
\bibitem [{\citenamefont {White}\ \emph {et~al.}(2023)\citenamefont {White}, \citenamefont {Opremcak}, \citenamefont {Sterling}, \citenamefont {Korotkov}, \citenamefont {Sank}, \citenamefont {Acharya}, \citenamefont {Ansmann}, \citenamefont {Arute}, \citenamefont {Arya}, \citenamefont {Bardin} \emph {et~al.}}]{white2023readout}%
  \BibitemOpen
  \bibfield  {author} {\bibinfo {author} {\bibfnamefont {T.~C.}\ \bibnamefont {White}}, \bibinfo {author} {\bibfnamefont {A.}~\bibnamefont {Opremcak}}, \bibinfo {author} {\bibfnamefont {G.}~\bibnamefont {Sterling}}, \bibinfo {author} {\bibfnamefont {A.}~\bibnamefont {Korotkov}}, \bibinfo {author} {\bibfnamefont {D.}~\bibnamefont {Sank}}, \bibinfo {author} {\bibfnamefont {R.}~\bibnamefont {Acharya}}, \bibinfo {author} {\bibfnamefont {M.}~\bibnamefont {Ansmann}}, \bibinfo {author} {\bibfnamefont {F.}~\bibnamefont {Arute}}, \bibinfo {author} {\bibfnamefont {K.}~\bibnamefont {Arya}}, \bibinfo {author} {\bibfnamefont {J.~C.}\ \bibnamefont {Bardin}}, \emph {et~al.},\ }\bibfield  {title} {\bibinfo {title} {Readout of a quantum processor with high dynamic range {Josephson} parametric amplifiers},\ }\href@noop {} {\bibfield  {journal} {\bibinfo  {journal} {Applied Physics Letters}\ }\textbf {\bibinfo {volume} {122}},\ \bibinfo {pages} {014001} (\bibinfo {year} {2023})}\BibitemShut {NoStop}%
\bibitem [{\citenamefont {Joshi}\ \emph {et~al.}(2025)\citenamefont {Joshi}, \citenamefont {Hazra}, \citenamefont {Ding}, \citenamefont {Miano}, \citenamefont {Dai}, \citenamefont {Umasankar}, \citenamefont {Kottandavida}, \citenamefont {Liu}, \citenamefont {Frunzio},\ and\ \citenamefont {Devoret}}]{joshi2025lumped}%
  \BibitemOpen
  \bibfield  {author} {\bibinfo {author} {\bibfnamefont {V.}~\bibnamefont {Joshi}}, \bibinfo {author} {\bibfnamefont {S.}~\bibnamefont {Hazra}}, \bibinfo {author} {\bibfnamefont {A.}~\bibnamefont {Ding}}, \bibinfo {author} {\bibfnamefont {A.}~\bibnamefont {Miano}}, \bibinfo {author} {\bibfnamefont {W.}~\bibnamefont {Dai}}, \bibinfo {author} {\bibfnamefont {G.}~\bibnamefont {Umasankar}}, \bibinfo {author} {\bibfnamefont {A.}~\bibnamefont {Kottandavida}}, \bibinfo {author} {\bibfnamefont {G.}~\bibnamefont {Liu}}, \bibinfo {author} {\bibfnamefont {L.}~\bibnamefont {Frunzio}},\ and\ \bibinfo {author} {\bibfnamefont {M.}~\bibnamefont {Devoret}},\ }\bibfield  {title} {\bibinfo {title} {Lumped-element broadband {SNAIL} parametric amplifier with on-chip pump filter for multiplexed readout},\ }\href@noop {} {\bibfield  {journal} {\bibinfo  {journal} {Physical Review Applied}\ }\textbf {\bibinfo {volume} {24}},\ \bibinfo {pages} {044003} (\bibinfo {year} {2025})}\BibitemShut {NoStop}%
\bibitem [{\citenamefont {Vijay}\ \emph {et~al.}(2011)\citenamefont {Vijay}, \citenamefont {Slichter},\ and\ \citenamefont {Siddiqi}}]{vijay2011observation}%
  \BibitemOpen
  \bibfield  {author} {\bibinfo {author} {\bibfnamefont {R.}~\bibnamefont {Vijay}}, \bibinfo {author} {\bibfnamefont {D.}~\bibnamefont {Slichter}},\ and\ \bibinfo {author} {\bibfnamefont {I.}~\bibnamefont {Siddiqi}},\ }\bibfield  {title} {\bibinfo {title} {Observation of quantum jumps in a superconducting artificial atom},\ }\href@noop {} {\bibfield  {journal} {\bibinfo  {journal} {Physical review letters}\ }\textbf {\bibinfo {volume} {106}},\ \bibinfo {pages} {110502} (\bibinfo {year} {2011})}\BibitemShut {NoStop}%
\bibitem [{\citenamefont {Hatridge}\ \emph {et~al.}(2011)\citenamefont {Hatridge}, \citenamefont {Vijay}, \citenamefont {Slichter}, \citenamefont {Clarke},\ and\ \citenamefont {Siddiqi}}]{hatridge2011dispersive}%
  \BibitemOpen
  \bibfield  {author} {\bibinfo {author} {\bibfnamefont {M.}~\bibnamefont {Hatridge}}, \bibinfo {author} {\bibfnamefont {R.}~\bibnamefont {Vijay}}, \bibinfo {author} {\bibfnamefont {D.}~\bibnamefont {Slichter}}, \bibinfo {author} {\bibfnamefont {J.}~\bibnamefont {Clarke}},\ and\ \bibinfo {author} {\bibfnamefont {I.}~\bibnamefont {Siddiqi}},\ }\bibfield  {title} {\bibinfo {title} {Dispersive magnetometry with a quantum limited {SQUID} parametric amplifier},\ }\href@noop {} {\bibfield  {journal} {\bibinfo  {journal} {Physical Review B—Condensed Matter and Materials Physics}\ }\textbf {\bibinfo {volume} {83}},\ \bibinfo {pages} {134501} (\bibinfo {year} {2011})}\BibitemShut {NoStop}%
\bibitem [{\citenamefont {Roy}\ and\ \citenamefont {Devoret}(2016)}]{roy2016introduction}%
  \BibitemOpen
  \bibfield  {author} {\bibinfo {author} {\bibfnamefont {A.}~\bibnamefont {Roy}}\ and\ \bibinfo {author} {\bibfnamefont {M.}~\bibnamefont {Devoret}},\ }\bibfield  {title} {\bibinfo {title} {Introduction to parametric amplification of quantum signals with {Josephson} circuits},\ }\href@noop {} {\bibfield  {journal} {\bibinfo  {journal} {Comptes Rendus Physique}\ }\textbf {\bibinfo {volume} {17}},\ \bibinfo {pages} {740} (\bibinfo {year} {2016})}\BibitemShut {NoStop}%
\bibitem [{\citenamefont {Roy}\ and\ \citenamefont {Devoret}(2018)}]{roy2018quantum}%
  \BibitemOpen
  \bibfield  {author} {\bibinfo {author} {\bibfnamefont {A.}~\bibnamefont {Roy}}\ and\ \bibinfo {author} {\bibfnamefont {M.}~\bibnamefont {Devoret}},\ }\bibfield  {title} {\bibinfo {title} {Quantum-limited parametric amplification with {Josephson} circuits in the regime of pump depletion},\ }\href@noop {} {\bibfield  {journal} {\bibinfo  {journal} {Physical Review B}\ }\textbf {\bibinfo {volume} {98}},\ \bibinfo {pages} {045405} (\bibinfo {year} {2018})}\BibitemShut {NoStop}%
\bibitem [{\citenamefont {Aumentado}(2020)}]{aumentado2020superconducting}%
  \BibitemOpen
  \bibfield  {author} {\bibinfo {author} {\bibfnamefont {J.}~\bibnamefont {Aumentado}},\ }\bibfield  {title} {\bibinfo {title} {Superconducting parametric amplifiers: The state of the art in {Josephson} parametric amplifiers},\ }\href@noop {} {\bibfield  {journal} {\bibinfo  {journal} {IEEE Microwave magazine}\ }\textbf {\bibinfo {volume} {21}},\ \bibinfo {pages} {45} (\bibinfo {year} {2020})}\BibitemShut {NoStop}%
\bibitem [{\citenamefont {Backes}\ \emph {et~al.}(2021)\citenamefont {Backes}, \citenamefont {Palken}, \citenamefont {Kenany}, \citenamefont {Brubaker}, \citenamefont {Cahn}, \citenamefont {Droster}, \citenamefont {Hilton}, \citenamefont {Ghosh}, \citenamefont {Jackson}, \citenamefont {Lamoreaux} \emph {et~al.}}]{backes2021quantum}%
  \BibitemOpen
  \bibfield  {author} {\bibinfo {author} {\bibfnamefont {K.~M.}\ \bibnamefont {Backes}}, \bibinfo {author} {\bibfnamefont {D.~A.}\ \bibnamefont {Palken}}, \bibinfo {author} {\bibfnamefont {S.~A.}\ \bibnamefont {Kenany}}, \bibinfo {author} {\bibfnamefont {B.~M.}\ \bibnamefont {Brubaker}}, \bibinfo {author} {\bibfnamefont {S.}~\bibnamefont {Cahn}}, \bibinfo {author} {\bibfnamefont {A.}~\bibnamefont {Droster}}, \bibinfo {author} {\bibfnamefont {G.~C.}\ \bibnamefont {Hilton}}, \bibinfo {author} {\bibfnamefont {S.}~\bibnamefont {Ghosh}}, \bibinfo {author} {\bibfnamefont {H.}~\bibnamefont {Jackson}}, \bibinfo {author} {\bibfnamefont {S.~K.}\ \bibnamefont {Lamoreaux}}, \emph {et~al.},\ }\bibfield  {title} {\bibinfo {title} {A quantum enhanced search for dark matter axions},\ }\href@noop {} {\bibfield  {journal} {\bibinfo  {journal} {Nature}\ }\textbf {\bibinfo {volume} {590}},\ \bibinfo {pages} {238} (\bibinfo {year} {2021})}\BibitemShut {NoStop}%
\bibitem [{\citenamefont {Uchaikin}\ \emph {et~al.}(2024)\citenamefont {Uchaikin}, \citenamefont {Kim}, \citenamefont {Kutlu}, \citenamefont {Ivanov}, \citenamefont {Kim}, \citenamefont {van Loo}, \citenamefont {Nakamura}, \citenamefont {Ahn}, \citenamefont {Oh}, \citenamefont {Ko},\ and\ \citenamefont {Semertzidis}}]{uchaikin2024josephson}%
  \BibitemOpen
  \bibfield  {author} {\bibinfo {author} {\bibfnamefont {S.~V.}\ \bibnamefont {Uchaikin}}, \bibinfo {author} {\bibfnamefont {J.}~\bibnamefont {Kim}}, \bibinfo {author} {\bibfnamefont {C.}~\bibnamefont {Kutlu}}, \bibinfo {author} {\bibfnamefont {B.~I.}\ \bibnamefont {Ivanov}}, \bibinfo {author} {\bibfnamefont {J.}~\bibnamefont {Kim}}, \bibinfo {author} {\bibfnamefont {A.~F.}\ \bibnamefont {van Loo}}, \bibinfo {author} {\bibfnamefont {Y.}~\bibnamefont {Nakamura}}, \bibinfo {author} {\bibfnamefont {S.}~\bibnamefont {Ahn}}, \bibinfo {author} {\bibfnamefont {S.}~\bibnamefont {Oh}}, \bibinfo {author} {\bibfnamefont {M.}~\bibnamefont {Ko}},\ and\ \bibinfo {author} {\bibfnamefont {Y.~K.}\ \bibnamefont {Semertzidis}},\ }\bibfield  {title} {\bibinfo {title} {{Josephson} parametric amplifier based quantum noise limited amplifier development for axion search experiments in {CAPP}},\ }\href@noop {} {\bibfield  {journal} {\bibinfo  {journal} {Frontiers in Physics}\ }\textbf {\bibinfo {volume} {12}},\ \bibinfo {pages}
  {1437680} (\bibinfo {year} {2024})}\BibitemShut {NoStop}%
\bibitem [{\citenamefont {Qing}\ \emph {et~al.}(2024)\citenamefont {Qing}, \citenamefont {Nguyen}, \citenamefont {Liu}, \citenamefont {Ren}, \citenamefont {Livingston}, \citenamefont {Goss}, \citenamefont {Hajr}, \citenamefont {Chistolini}, \citenamefont {Pedramrazi}, \citenamefont {Santiago}, \citenamefont {Luo},\ and\ \citenamefont {Siddiqi}}]{ranzani2024broadband}%
  \BibitemOpen
  \bibfield  {author} {\bibinfo {author} {\bibfnamefont {B.}~\bibnamefont {Qing}}, \bibinfo {author} {\bibfnamefont {L.~B.}\ \bibnamefont {Nguyen}}, \bibinfo {author} {\bibfnamefont {X.}~\bibnamefont {Liu}}, \bibinfo {author} {\bibfnamefont {H.}~\bibnamefont {Ren}}, \bibinfo {author} {\bibfnamefont {W.~P.}\ \bibnamefont {Livingston}}, \bibinfo {author} {\bibfnamefont {N.}~\bibnamefont {Goss}}, \bibinfo {author} {\bibfnamefont {A.}~\bibnamefont {Hajr}}, \bibinfo {author} {\bibfnamefont {T.}~\bibnamefont {Chistolini}}, \bibinfo {author} {\bibfnamefont {Z.}~\bibnamefont {Pedramrazi}}, \bibinfo {author} {\bibfnamefont {D.~I.}\ \bibnamefont {Santiago}}, \bibinfo {author} {\bibfnamefont {J.}~\bibnamefont {Luo}},\ and\ \bibinfo {author} {\bibfnamefont {I.}~\bibnamefont {Siddiqi}},\ }\bibfield  {title} {\bibinfo {title} {Broadband coplanar-waveguide-based impedance-transformed {Josephson} parametric amplifier},\ }\href {https://doi.org/10.1103/PhysRevResearch.6.L012035} {\bibfield  {journal} {\bibinfo  {journal}
  {Physical Review Research}\ }\textbf {\bibinfo {volume} {6}},\ \bibinfo {pages} {L012035} (\bibinfo {year} {2024})}\BibitemShut {NoStop}%
\bibitem [{\citenamefont {Yurke}\ \emph {et~al.}(1988)\citenamefont {Yurke}, \citenamefont {Kaminsky}, \citenamefont {Miller}, \citenamefont {Whittaker}, \citenamefont {Smith}, \citenamefont {Silver},\ and\ \citenamefont {Simon}}]{yurke1988observation}%
  \BibitemOpen
  \bibfield  {author} {\bibinfo {author} {\bibfnamefont {B.}~\bibnamefont {Yurke}}, \bibinfo {author} {\bibfnamefont {P.}~\bibnamefont {Kaminsky}}, \bibinfo {author} {\bibfnamefont {R.}~\bibnamefont {Miller}}, \bibinfo {author} {\bibfnamefont {E.}~\bibnamefont {Whittaker}}, \bibinfo {author} {\bibfnamefont {A.}~\bibnamefont {Smith}}, \bibinfo {author} {\bibfnamefont {A.}~\bibnamefont {Silver}},\ and\ \bibinfo {author} {\bibfnamefont {R.}~\bibnamefont {Simon}},\ }\bibfield  {title} {\bibinfo {title} {Observation of 4.2-k equilibrium-noise squeezing via a {Josephson}-parametric amplifier},\ }\href@noop {} {\bibfield  {journal} {\bibinfo  {journal} {Physical review letters}\ }\textbf {\bibinfo {volume} {60}},\ \bibinfo {pages} {764} (\bibinfo {year} {1988})}\BibitemShut {NoStop}%
\bibitem [{\citenamefont {Yurke}\ \emph {et~al.}(1989)\citenamefont {Yurke}, \citenamefont {Corruccini}, \citenamefont {Kaminsky}, \citenamefont {Rupp}, \citenamefont {Smith}, \citenamefont {Silver}, \citenamefont {Simon},\ and\ \citenamefont {Whittaker}}]{yurke1989observation}%
  \BibitemOpen
  \bibfield  {author} {\bibinfo {author} {\bibfnamefont {B.}~\bibnamefont {Yurke}}, \bibinfo {author} {\bibfnamefont {L.}~\bibnamefont {Corruccini}}, \bibinfo {author} {\bibfnamefont {P.}~\bibnamefont {Kaminsky}}, \bibinfo {author} {\bibfnamefont {L.}~\bibnamefont {Rupp}}, \bibinfo {author} {\bibfnamefont {A.}~\bibnamefont {Smith}}, \bibinfo {author} {\bibfnamefont {A.}~\bibnamefont {Silver}}, \bibinfo {author} {\bibfnamefont {R.}~\bibnamefont {Simon}},\ and\ \bibinfo {author} {\bibfnamefont {E.}~\bibnamefont {Whittaker}},\ }\bibfield  {title} {\bibinfo {title} {Observation of parametric amplification and deamplification in a {Josephson} parametric amplifier},\ }\href@noop {} {\bibfield  {journal} {\bibinfo  {journal} {Physical Review A}\ }\textbf {\bibinfo {volume} {39}},\ \bibinfo {pages} {2519} (\bibinfo {year} {1989})}\BibitemShut {NoStop}%
\bibitem [{\citenamefont {Kaufman}\ \emph {et~al.}(2025)\citenamefont {Kaufman}, \citenamefont {Liu}, \citenamefont {Cicak}, \citenamefont {Mesits}, \citenamefont {Xia}, \citenamefont {Zhou}, \citenamefont {Nowicki}, \citenamefont {Aumentado}, \citenamefont {Pekker},\ and\ \citenamefont {Hatridge}}]{kaufman2025simple}%
  \BibitemOpen
  \bibfield  {author} {\bibinfo {author} {\bibfnamefont {R.}~\bibnamefont {Kaufman}}, \bibinfo {author} {\bibfnamefont {C.}~\bibnamefont {Liu}}, \bibinfo {author} {\bibfnamefont {K.}~\bibnamefont {Cicak}}, \bibinfo {author} {\bibfnamefont {B.}~\bibnamefont {Mesits}}, \bibinfo {author} {\bibfnamefont {M.}~\bibnamefont {Xia}}, \bibinfo {author} {\bibfnamefont {C.}~\bibnamefont {Zhou}}, \bibinfo {author} {\bibfnamefont {M.}~\bibnamefont {Nowicki}}, \bibinfo {author} {\bibfnamefont {J.}~\bibnamefont {Aumentado}}, \bibinfo {author} {\bibfnamefont {D.}~\bibnamefont {Pekker}},\ and\ \bibinfo {author} {\bibfnamefont {M.}~\bibnamefont {Hatridge}},\ }\bibfield  {title} {\bibinfo {title} {Simple high-saturation-power quantum-limited rf-{SQUID}-array-based {Josephson} parametric amplifiers},\ }\href@noop {} {\bibfield  {journal} {\bibinfo  {journal} {Physical Review Applied}\ }\textbf {\bibinfo {volume} {24}},\ \bibinfo {pages} {014052} (\bibinfo {year} {2025})}\BibitemShut {NoStop}%
\bibitem [{\citenamefont {Mutus}\ \emph {et~al.}(2014)\citenamefont {Mutus}, \citenamefont {White}, \citenamefont {Barends}, \citenamefont {Chen}, \citenamefont {Chen}, \citenamefont {Chiaro}, \citenamefont {Dunsworth}, \citenamefont {Jeffrey}, \citenamefont {Kelly}, \citenamefont {Megrant}, \citenamefont {Neill}, \citenamefont {O'Malley}, \citenamefont {Roushan}, \citenamefont {Sank}, \citenamefont {Vainsencher}, \citenamefont {Wenner}, \citenamefont {Sundqvist}, \citenamefont {Cleland},\ and\ \citenamefont {Martinis}}]{mutus2014strong}%
  \BibitemOpen
  \bibfield  {author} {\bibinfo {author} {\bibfnamefont {J.~Y.}\ \bibnamefont {Mutus}}, \bibinfo {author} {\bibfnamefont {T.~C.}\ \bibnamefont {White}}, \bibinfo {author} {\bibfnamefont {R.}~\bibnamefont {Barends}}, \bibinfo {author} {\bibfnamefont {Y.}~\bibnamefont {Chen}}, \bibinfo {author} {\bibfnamefont {Z.}~\bibnamefont {Chen}}, \bibinfo {author} {\bibfnamefont {B.}~\bibnamefont {Chiaro}}, \bibinfo {author} {\bibfnamefont {A.}~\bibnamefont {Dunsworth}}, \bibinfo {author} {\bibfnamefont {E.}~\bibnamefont {Jeffrey}}, \bibinfo {author} {\bibfnamefont {J.}~\bibnamefont {Kelly}}, \bibinfo {author} {\bibfnamefont {A.}~\bibnamefont {Megrant}}, \bibinfo {author} {\bibfnamefont {C.}~\bibnamefont {Neill}}, \bibinfo {author} {\bibfnamefont {P.~J.~J.}\ \bibnamefont {O'Malley}}, \bibinfo {author} {\bibfnamefont {P.}~\bibnamefont {Roushan}}, \bibinfo {author} {\bibfnamefont {D.}~\bibnamefont {Sank}}, \bibinfo {author} {\bibfnamefont {A.}~\bibnamefont {Vainsencher}}, \bibinfo {author} {\bibfnamefont {J.}~\bibnamefont
  {Wenner}}, \bibinfo {author} {\bibfnamefont {K.~M.}\ \bibnamefont {Sundqvist}}, \bibinfo {author} {\bibfnamefont {A.~N.}\ \bibnamefont {Cleland}},\ and\ \bibinfo {author} {\bibfnamefont {J.~M.}\ \bibnamefont {Martinis}},\ }\bibfield  {title} {\bibinfo {title} {Strong environmental coupling in a {Josephson} parametric amplifier},\ }\href@noop {} {\bibfield  {journal} {\bibinfo  {journal} {Applied Physics Letters}\ }\textbf {\bibinfo {volume} {104}},\ \bibinfo {pages} {263513} (\bibinfo {year} {2014})}\BibitemShut {NoStop}%
\bibitem [{\citenamefont {Ezenkova}\ \emph {et~al.}(2022)\citenamefont {Ezenkova}, \citenamefont {Moskalev}, \citenamefont {Smirnov}, \citenamefont {Ivanov}, \citenamefont {Matanin}, \citenamefont {Polozov}, \citenamefont {Echeistov}, \citenamefont {Malevannaya}, \citenamefont {Samoylov}, \citenamefont {Zikiy},\ and\ \citenamefont {Rodionov}}]{ezenkova2022broadband}%
  \BibitemOpen
  \bibfield  {author} {\bibinfo {author} {\bibfnamefont {D.}~\bibnamefont {Ezenkova}}, \bibinfo {author} {\bibfnamefont {D.}~\bibnamefont {Moskalev}}, \bibinfo {author} {\bibfnamefont {N.}~\bibnamefont {Smirnov}}, \bibinfo {author} {\bibfnamefont {A.}~\bibnamefont {Ivanov}}, \bibinfo {author} {\bibfnamefont {A.}~\bibnamefont {Matanin}}, \bibinfo {author} {\bibfnamefont {V.}~\bibnamefont {Polozov}}, \bibinfo {author} {\bibfnamefont {V.}~\bibnamefont {Echeistov}}, \bibinfo {author} {\bibfnamefont {E.}~\bibnamefont {Malevannaya}}, \bibinfo {author} {\bibfnamefont {A.}~\bibnamefont {Samoylov}}, \bibinfo {author} {\bibfnamefont {E.}~\bibnamefont {Zikiy}},\ and\ \bibinfo {author} {\bibfnamefont {I.}~\bibnamefont {Rodionov}},\ }\bibfield  {title} {\bibinfo {title} {Broadband {SNAIL} parametric amplifier with microstrip impedance transformer},\ }\href@noop {} {\bibfield  {journal} {\bibinfo  {journal} {Applied Physics Letters}\ }\textbf {\bibinfo {volume} {121}},\ \bibinfo {pages} {232601} (\bibinfo {year}
  {2022})}\BibitemShut {NoStop}%
\bibitem [{\citenamefont {Kaufman}\ \emph {et~al.}(2023)\citenamefont {Kaufman}, \citenamefont {White}, \citenamefont {Dykman}, \citenamefont {Iorio}, \citenamefont {Sterling}, \citenamefont {Hong}, \citenamefont {Opremcak}, \citenamefont {Bengtsson}, \citenamefont {Faoro}, \citenamefont {Bardin}, \citenamefont {Burger}, \citenamefont {Gasca},\ and\ \citenamefont {Naaman}}]{kaufman2023josephson}%
  \BibitemOpen
  \bibfield  {author} {\bibinfo {author} {\bibfnamefont {R.}~\bibnamefont {Kaufman}}, \bibinfo {author} {\bibfnamefont {T.}~\bibnamefont {White}}, \bibinfo {author} {\bibfnamefont {M.~I.}\ \bibnamefont {Dykman}}, \bibinfo {author} {\bibfnamefont {A.}~\bibnamefont {Iorio}}, \bibinfo {author} {\bibfnamefont {G.}~\bibnamefont {Sterling}}, \bibinfo {author} {\bibfnamefont {S.}~\bibnamefont {Hong}}, \bibinfo {author} {\bibfnamefont {A.}~\bibnamefont {Opremcak}}, \bibinfo {author} {\bibfnamefont {A.}~\bibnamefont {Bengtsson}}, \bibinfo {author} {\bibfnamefont {L.}~\bibnamefont {Faoro}}, \bibinfo {author} {\bibfnamefont {J.~C.}\ \bibnamefont {Bardin}}, \bibinfo {author} {\bibfnamefont {T.}~\bibnamefont {Burger}}, \bibinfo {author} {\bibfnamefont {R.}~\bibnamefont {Gasca}},\ and\ \bibinfo {author} {\bibfnamefont {O.}~\bibnamefont {Naaman}},\ }\bibfield  {title} {\bibinfo {title} {{Josephson} parametric amplifier with {Chebyshev} gain profile and high saturation},\ }\href@noop {} {\bibfield  {journal} {\bibinfo
  {journal} {Physical Review Applied}\ }\textbf {\bibinfo {volume} {20}},\ \bibinfo {pages} {054058} (\bibinfo {year} {2023})}\BibitemShut {NoStop}%
\bibitem [{\citenamefont {O'Brien}\ \emph {et~al.}(2014)\citenamefont {O'Brien}, \citenamefont {Macklin}, \citenamefont {Siddiqi},\ and\ \citenamefont {Zhang}}]{obrien2014resonant}%
  \BibitemOpen
  \bibfield  {author} {\bibinfo {author} {\bibfnamefont {K.}~\bibnamefont {O'Brien}}, \bibinfo {author} {\bibfnamefont {C.}~\bibnamefont {Macklin}}, \bibinfo {author} {\bibfnamefont {I.}~\bibnamefont {Siddiqi}},\ and\ \bibinfo {author} {\bibfnamefont {X.}~\bibnamefont {Zhang}},\ }\bibfield  {title} {\bibinfo {title} {Resonant phase matching of {Josephson} junction traveling wave parametric amplifiers},\ }\href@noop {} {\bibfield  {journal} {\bibinfo  {journal} {Physical Review Letters}\ }\textbf {\bibinfo {volume} {113}},\ \bibinfo {pages} {157001} (\bibinfo {year} {2014})}\BibitemShut {NoStop}%
\bibitem [{\citenamefont {Macklin}\ \emph {et~al.}(2015)\citenamefont {Macklin}, \citenamefont {O'Brien}, \citenamefont {Hover}, \citenamefont {Schwartz}, \citenamefont {Bolkhovsky}, \citenamefont {Zhang}, \citenamefont {Oliver},\ and\ \citenamefont {Siddiqi}}]{macklin2015near}%
  \BibitemOpen
  \bibfield  {author} {\bibinfo {author} {\bibfnamefont {C.}~\bibnamefont {Macklin}}, \bibinfo {author} {\bibfnamefont {K.}~\bibnamefont {O'Brien}}, \bibinfo {author} {\bibfnamefont {D.}~\bibnamefont {Hover}}, \bibinfo {author} {\bibfnamefont {M.~E.}\ \bibnamefont {Schwartz}}, \bibinfo {author} {\bibfnamefont {V.}~\bibnamefont {Bolkhovsky}}, \bibinfo {author} {\bibfnamefont {X.}~\bibnamefont {Zhang}}, \bibinfo {author} {\bibfnamefont {W.~D.}\ \bibnamefont {Oliver}},\ and\ \bibinfo {author} {\bibfnamefont {I.}~\bibnamefont {Siddiqi}},\ }\bibfield  {title} {\bibinfo {title} {A near--quantum-limited {Josephson} traveling-wave parametric amplifier},\ }\href@noop {} {\bibfield  {journal} {\bibinfo  {journal} {Science}\ }\textbf {\bibinfo {volume} {350}},\ \bibinfo {pages} {307} (\bibinfo {year} {2015})}\BibitemShut {NoStop}%
\bibitem [{\citenamefont {White}\ \emph {et~al.}(2015)\citenamefont {White}, \citenamefont {Mutus}, \citenamefont {Dresselhaus}, \citenamefont {Kelly}, \citenamefont {Barends}, \citenamefont {Jeffrey}, \citenamefont {Sank}, \citenamefont {Megrant}, \citenamefont {Neill}, \citenamefont {O'Malley}, \citenamefont {Roushan}, \citenamefont {Vainsencher}, \citenamefont {Wenner}, \citenamefont {Korotkov}, \citenamefont {Cleland},\ and\ \citenamefont {Martinis}}]{white2015traveling}%
  \BibitemOpen
  \bibfield  {author} {\bibinfo {author} {\bibfnamefont {T.~C.}\ \bibnamefont {White}}, \bibinfo {author} {\bibfnamefont {J.~Y.}\ \bibnamefont {Mutus}}, \bibinfo {author} {\bibfnamefont {P.~D.}\ \bibnamefont {Dresselhaus}}, \bibinfo {author} {\bibfnamefont {J.}~\bibnamefont {Kelly}}, \bibinfo {author} {\bibfnamefont {R.}~\bibnamefont {Barends}}, \bibinfo {author} {\bibfnamefont {E.}~\bibnamefont {Jeffrey}}, \bibinfo {author} {\bibfnamefont {D.}~\bibnamefont {Sank}}, \bibinfo {author} {\bibfnamefont {A.}~\bibnamefont {Megrant}}, \bibinfo {author} {\bibfnamefont {C.}~\bibnamefont {Neill}}, \bibinfo {author} {\bibfnamefont {P.~J.~J.}\ \bibnamefont {O'Malley}}, \bibinfo {author} {\bibfnamefont {P.}~\bibnamefont {Roushan}}, \bibinfo {author} {\bibfnamefont {A.}~\bibnamefont {Vainsencher}}, \bibinfo {author} {\bibfnamefont {J.}~\bibnamefont {Wenner}}, \bibinfo {author} {\bibfnamefont {A.~N.}\ \bibnamefont {Korotkov}}, \bibinfo {author} {\bibfnamefont {A.~N.}\ \bibnamefont {Cleland}},\ and\ \bibinfo {author}
  {\bibfnamefont {J.~M.}\ \bibnamefont {Martinis}},\ }\bibfield  {title} {\bibinfo {title} {Traveling wave parametric amplifier with {Josephson} junctions using minimal resonator phase matching},\ }\href@noop {} {\bibfield  {journal} {\bibinfo  {journal} {Applied Physics Letters}\ }\textbf {\bibinfo {volume} {106}},\ \bibinfo {pages} {242601} (\bibinfo {year} {2015})}\BibitemShut {NoStop}%
\bibitem [{\citenamefont {Planat}\ \emph {et~al.}(2020)\citenamefont {Planat}, \citenamefont {Ranadive}, \citenamefont {Dassonneville}, \citenamefont {Puertas}, \citenamefont {Leger}, \citenamefont {Naud}, \citenamefont {Buisson}, \citenamefont {Hasch-Guichard}, \citenamefont {Basko},\ and\ \citenamefont {Roch}}]{planat2020photonic}%
  \BibitemOpen
  \bibfield  {author} {\bibinfo {author} {\bibfnamefont {L.}~\bibnamefont {Planat}}, \bibinfo {author} {\bibfnamefont {A.}~\bibnamefont {Ranadive}}, \bibinfo {author} {\bibfnamefont {R.}~\bibnamefont {Dassonneville}}, \bibinfo {author} {\bibfnamefont {G.}~\bibnamefont {Puertas}}, \bibinfo {author} {\bibfnamefont {S.}~\bibnamefont {Leger}}, \bibinfo {author} {\bibfnamefont {C.}~\bibnamefont {Naud}}, \bibinfo {author} {\bibfnamefont {O.}~\bibnamefont {Buisson}}, \bibinfo {author} {\bibfnamefont {W.}~\bibnamefont {Hasch-Guichard}}, \bibinfo {author} {\bibfnamefont {D.~M.}\ \bibnamefont {Basko}},\ and\ \bibinfo {author} {\bibfnamefont {N.}~\bibnamefont {Roch}},\ }\bibfield  {title} {\bibinfo {title} {Photonic-crystal {Josephson} traveling-wave parametric amplifier},\ }\href@noop {} {\bibfield  {journal} {\bibinfo  {journal} {Physical Review X}\ }\textbf {\bibinfo {volume} {10}},\ \bibinfo {pages} {021021} (\bibinfo {year} {2020})}\BibitemShut {NoStop}%
\bibitem [{\citenamefont {Zorin}\ \emph {et~al.}(2017)\citenamefont {Zorin}, \citenamefont {Khabipov}, \citenamefont {Dietel},\ and\ \citenamefont {Dolata}}]{zorin2017traveling}%
  \BibitemOpen
  \bibfield  {author} {\bibinfo {author} {\bibfnamefont {A.~B.}\ \bibnamefont {Zorin}}, \bibinfo {author} {\bibfnamefont {M.}~\bibnamefont {Khabipov}}, \bibinfo {author} {\bibfnamefont {J.}~\bibnamefont {Dietel}},\ and\ \bibinfo {author} {\bibfnamefont {R.}~\bibnamefont {Dolata}},\ }\bibfield  {title} {\bibinfo {title} {Traveling-wave parametric amplifier based on three-wave mixing in a {Josephson} metamaterial},\ }in\ \href@noop {} {\emph {\bibinfo {booktitle} {2017 16th International Superconductive Electronics Conference (ISEC)}}}\ (\bibinfo {organization} {IEEE},\ \bibinfo {year} {2017})\ pp.\ \bibinfo {pages} {1--3}\BibitemShut {NoStop}%
\bibitem [{\citenamefont {{\'O}~Peat{\'a}in}\ \emph {et~al.}(2023)\citenamefont {{\'O}~Peat{\'a}in}, \citenamefont {Dixon}, \citenamefont {Meeson}, \citenamefont {Williams}, \citenamefont {Kafanov},\ and\ \citenamefont {Pashkin}}]{o2023simulating}%
  \BibitemOpen
  \bibfield  {author} {\bibinfo {author} {\bibfnamefont {S.}~\bibnamefont {{\'O}~Peat{\'a}in}}, \bibinfo {author} {\bibfnamefont {T.}~\bibnamefont {Dixon}}, \bibinfo {author} {\bibfnamefont {P.~J.}\ \bibnamefont {Meeson}}, \bibinfo {author} {\bibfnamefont {J.}~\bibnamefont {Williams}}, \bibinfo {author} {\bibfnamefont {S.}~\bibnamefont {Kafanov}},\ and\ \bibinfo {author} {\bibfnamefont {Y.~A.}\ \bibnamefont {Pashkin}},\ }\bibfield  {title} {\bibinfo {title} {Simulating the effects of fabrication tolerance on the performance of {Josephson} travelling wave parametric amplifiers},\ }\href@noop {} {\bibfield  {journal} {\bibinfo  {journal} {Superconductor Science and Technology}\ }\textbf {\bibinfo {volume} {36}},\ \bibinfo {pages} {045017} (\bibinfo {year} {2023})}\BibitemShut {NoStop}%
\bibitem [{\citenamefont {Zhao}\ \emph {et~al.}(2019)\citenamefont {Zhao}, \citenamefont {Withington}, \citenamefont {Goldie},\ and\ \citenamefont {Thomas}}]{zhao2019loss}%
  \BibitemOpen
  \bibfield  {author} {\bibinfo {author} {\bibfnamefont {S.}~\bibnamefont {Zhao}}, \bibinfo {author} {\bibfnamefont {S.}~\bibnamefont {Withington}}, \bibinfo {author} {\bibfnamefont {D.~J.}\ \bibnamefont {Goldie}},\ and\ \bibinfo {author} {\bibfnamefont {C.~N.}\ \bibnamefont {Thomas}},\ }\bibfield  {title} {\bibinfo {title} {Loss and saturation in superconducting travelling-wave parametric amplifiers},\ }\href@noop {} {\bibfield  {journal} {\bibinfo  {journal} {Journal of Physics D: Applied Physics}\ }\textbf {\bibinfo {volume} {52}},\ \bibinfo {pages} {415301} (\bibinfo {year} {2019})}\BibitemShut {NoStop}%
\bibitem [{\citenamefont {Chang}\ \emph {et~al.}(2025)\citenamefont {Chang}, \citenamefont {van Loo}, \citenamefont {Hung}, \citenamefont {Zhou}, \citenamefont {Tamate},\ and\ \citenamefont {Nakamura}}]{chang2025josephson}%
  \BibitemOpen
  \bibfield  {author} {\bibinfo {author} {\bibfnamefont {C.~W.~S.}\ \bibnamefont {Chang}}, \bibinfo {author} {\bibfnamefont {A.~F.}\ \bibnamefont {van Loo}}, \bibinfo {author} {\bibfnamefont {C.-C.}\ \bibnamefont {Hung}}, \bibinfo {author} {\bibfnamefont {Y.}~\bibnamefont {Zhou}}, \bibinfo {author} {\bibfnamefont {S.}~\bibnamefont {Tamate}},\ and\ \bibinfo {author} {\bibfnamefont {Y.}~\bibnamefont {Nakamura}},\ }\bibfield  {title} {\bibinfo {title} {{Josephson} traveling-wave parametric amplifier based on low-intrinsic-loss coplanar lumped-element waveguide},\ }\href@noop {} {\bibfield  {journal} {\bibinfo  {journal} {Physical Review Applied}\ }\textbf {\bibinfo {volume} {24}},\ \bibinfo {pages} {044081} (\bibinfo {year} {2025})}\BibitemShut {NoStop}%
\bibitem [{\citenamefont {Naaman}\ and\ \citenamefont {Aumentado}(2022)}]{naaman2022synthesis}%
  \BibitemOpen
  \bibfield  {author} {\bibinfo {author} {\bibfnamefont {O.}~\bibnamefont {Naaman}}\ and\ \bibinfo {author} {\bibfnamefont {J.}~\bibnamefont {Aumentado}},\ }\bibfield  {title} {\bibinfo {title} {Synthesis of parametrically coupled networks},\ }\href@noop {} {\bibfield  {journal} {\bibinfo  {journal} {PRX Quantum}\ }\textbf {\bibinfo {volume} {3}},\ \bibinfo {pages} {020201} (\bibinfo {year} {2022})}\BibitemShut {NoStop}%
\bibitem [{\citenamefont {Moskaleva}\ \emph {et~al.}(2024)\citenamefont {Moskaleva}, \citenamefont {Smirnov}, \citenamefont {Moskalev}, \citenamefont {Ivanov}, \citenamefont {Matanin}, \citenamefont {Baklykov}, \citenamefont {Teleganov}, \citenamefont {Polozov}, \citenamefont {Echeistov}, \citenamefont {Malevannaya} \emph {et~al.}}]{moskaleva2024lumped}%
  \BibitemOpen
  \bibfield  {author} {\bibinfo {author} {\bibfnamefont {D.}~\bibnamefont {Moskaleva}}, \bibinfo {author} {\bibfnamefont {N.}~\bibnamefont {Smirnov}}, \bibinfo {author} {\bibfnamefont {D.}~\bibnamefont {Moskalev}}, \bibinfo {author} {\bibfnamefont {A.}~\bibnamefont {Ivanov}}, \bibinfo {author} {\bibfnamefont {A.}~\bibnamefont {Matanin}}, \bibinfo {author} {\bibfnamefont {D.}~\bibnamefont {Baklykov}}, \bibinfo {author} {\bibfnamefont {M.}~\bibnamefont {Teleganov}}, \bibinfo {author} {\bibfnamefont {V.}~\bibnamefont {Polozov}}, \bibinfo {author} {\bibfnamefont {V.}~\bibnamefont {Echeistov}}, \bibinfo {author} {\bibfnamefont {E.}~\bibnamefont {Malevannaya}}, \emph {et~al.},\ }\bibfield  {title} {\bibinfo {title} {Lumped-element {SNAIL} parametric amplifier with two-pole matching network},\ }\href@noop {} {\bibfield  {journal} {\bibinfo  {journal} {Applied Physics Letters}\ }\textbf {\bibinfo {volume} {125}},\ \bibinfo {pages} {164003} (\bibinfo {year} {2024})}\BibitemShut {NoStop}%
\bibitem [{\citenamefont {Frattini}\ \emph {et~al.}(2017)\citenamefont {Frattini}, \citenamefont {Vool}, \citenamefont {Shankar}, \citenamefont {Narla}, \citenamefont {Sliwa},\ and\ \citenamefont {Devoret}}]{frattini2017wave}%
  \BibitemOpen
  \bibfield  {author} {\bibinfo {author} {\bibfnamefont {N.~E.}\ \bibnamefont {Frattini}}, \bibinfo {author} {\bibfnamefont {U.}~\bibnamefont {Vool}}, \bibinfo {author} {\bibfnamefont {S.}~\bibnamefont {Shankar}}, \bibinfo {author} {\bibfnamefont {A.}~\bibnamefont {Narla}}, \bibinfo {author} {\bibfnamefont {K.~M.}\ \bibnamefont {Sliwa}},\ and\ \bibinfo {author} {\bibfnamefont {M.~H.}\ \bibnamefont {Devoret}},\ }\bibfield  {title} {\bibinfo {title} {3-wave mixing {Josephson} dipole element},\ }\href@noop {} {\bibfield  {journal} {\bibinfo  {journal} {Applied Physics Letters}\ }\textbf {\bibinfo {volume} {110}},\ \bibinfo {pages} {222603} (\bibinfo {year} {2017})}\BibitemShut {NoStop}%
\bibitem [{\citenamefont {Frattini}\ \emph {et~al.}(2018)\citenamefont {Frattini}, \citenamefont {Sivak}, \citenamefont {Lingenfelter}, \citenamefont {Shankar},\ and\ \citenamefont {Devoret}}]{frattini2018optimizing}%
  \BibitemOpen
  \bibfield  {author} {\bibinfo {author} {\bibfnamefont {N.~E.}\ \bibnamefont {Frattini}}, \bibinfo {author} {\bibfnamefont {V.~V.}\ \bibnamefont {Sivak}}, \bibinfo {author} {\bibfnamefont {A.}~\bibnamefont {Lingenfelter}}, \bibinfo {author} {\bibfnamefont {S.}~\bibnamefont {Shankar}},\ and\ \bibinfo {author} {\bibfnamefont {M.~H.}\ \bibnamefont {Devoret}},\ }\bibfield  {title} {\bibinfo {title} {Optimizing the nonlinearity and dissipation of a {SNAIL} parametric amplifier for dynamic range},\ }\href@noop {} {\bibfield  {journal} {\bibinfo  {journal} {Physical Review Applied}\ }\textbf {\bibinfo {volume} {10}},\ \bibinfo {pages} {054020} (\bibinfo {year} {2018})}\BibitemShut {NoStop}%
\bibitem [{\citenamefont {Sivak}\ \emph {et~al.}(2019)\citenamefont {Sivak}, \citenamefont {Frattini}, \citenamefont {Joshi}, \citenamefont {Lingenfelter}, \citenamefont {Shankar},\ and\ \citenamefont {Devoret}}]{sivak2019kerr}%
  \BibitemOpen
  \bibfield  {author} {\bibinfo {author} {\bibfnamefont {V.~V.}\ \bibnamefont {Sivak}}, \bibinfo {author} {\bibfnamefont {N.~E.}\ \bibnamefont {Frattini}}, \bibinfo {author} {\bibfnamefont {V.~R.}\ \bibnamefont {Joshi}}, \bibinfo {author} {\bibfnamefont {A.}~\bibnamefont {Lingenfelter}}, \bibinfo {author} {\bibfnamefont {S.}~\bibnamefont {Shankar}},\ and\ \bibinfo {author} {\bibfnamefont {M.~H.}\ \bibnamefont {Devoret}},\ }\bibfield  {title} {\bibinfo {title} {Kerr-free three-wave mixing in superconducting quantum circuits},\ }\href@noop {} {\bibfield  {journal} {\bibinfo  {journal} {Physical Review Applied}\ }\textbf {\bibinfo {volume} {11}},\ \bibinfo {pages} {054060} (\bibinfo {year} {2019})}\BibitemShut {NoStop}%
\bibitem [{\citenamefont {Ketchen}\ \emph {et~al.}(1985)\citenamefont {Ketchen}, \citenamefont {Gallagher}, \citenamefont {Kleinsasser}, \citenamefont {Murphy},\ and\ \citenamefont {Clem}}]{ketchen1985dc}%
  \BibitemOpen
  \bibfield  {author} {\bibinfo {author} {\bibfnamefont {M.~B.}\ \bibnamefont {Ketchen}}, \bibinfo {author} {\bibfnamefont {W.~J.}\ \bibnamefont {Gallagher}}, \bibinfo {author} {\bibfnamefont {A.~W.}\ \bibnamefont {Kleinsasser}}, \bibinfo {author} {\bibfnamefont {S.}~\bibnamefont {Murphy}},\ and\ \bibinfo {author} {\bibfnamefont {J.~R.}\ \bibnamefont {Clem}},\ }\bibfield  {title} {\bibinfo {title} {dc {SQUID} flux focuser},\ }in\ \href@noop {} {\emph {\bibinfo {booktitle} {{SQUID} '85: Superconducting Quantum Interference Devices and their Applications}}},\ \bibinfo {editor} {edited by\ \bibinfo {editor} {\bibfnamefont {H.~D.}\ \bibnamefont {Hahlbohm}}\ and\ \bibinfo {editor} {\bibfnamefont {H.}~\bibnamefont {L{\"u}bbig}}}\ (\bibinfo  {publisher} {Walter de Gruyter},\ \bibinfo {year} {1985})\ p.\ \bibinfo {pages} {865}\BibitemShut {NoStop}%
\bibitem [{\citenamefont {Miklich}\ \emph {et~al.}(1991)\citenamefont {Miklich}, \citenamefont {Kingston}, \citenamefont {Wellstood}, \citenamefont {Clarke}, \citenamefont {Colclough}, \citenamefont {Char},\ and\ \citenamefont {Zaharchuk}}]{miklich1991sensitive}%
  \BibitemOpen
  \bibfield  {author} {\bibinfo {author} {\bibfnamefont {A.~H.}\ \bibnamefont {Miklich}}, \bibinfo {author} {\bibfnamefont {J.~J.}\ \bibnamefont {Kingston}}, \bibinfo {author} {\bibfnamefont {F.~C.}\ \bibnamefont {Wellstood}}, \bibinfo {author} {\bibfnamefont {J.}~\bibnamefont {Clarke}}, \bibinfo {author} {\bibfnamefont {M.~S.}\ \bibnamefont {Colclough}}, \bibinfo {author} {\bibfnamefont {K.}~\bibnamefont {Char}},\ and\ \bibinfo {author} {\bibfnamefont {G.}~\bibnamefont {Zaharchuk}},\ }\bibfield  {title} {\bibinfo {title} {Sensitive {YBa2Cu3O7-x} thin-film magnetometer},\ }\href@noop {} {\bibfield  {journal} {\bibinfo  {journal} {Applied Physics Letters}\ }\textbf {\bibinfo {volume} {59}},\ \bibinfo {pages} {988} (\bibinfo {year} {1991})}\BibitemShut {NoStop}%
\bibitem [{\citenamefont {Chesca}\ \emph {et~al.}(2015)\citenamefont {Chesca}, \citenamefont {John},\ and\ \citenamefont {Mellor}}]{chesca2015flux}%
  \BibitemOpen
  \bibfield  {author} {\bibinfo {author} {\bibfnamefont {B.}~\bibnamefont {Chesca}}, \bibinfo {author} {\bibfnamefont {D.}~\bibnamefont {John}},\ and\ \bibinfo {author} {\bibfnamefont {C.~J.}\ \bibnamefont {Mellor}},\ }\bibfield  {title} {\bibinfo {title} {Flux-coherent series {SQUID} array magnetometers operating above 77 k with superior white flux noise than single-{SQUID}s at 4.2 k},\ }\href@noop {} {\bibfield  {journal} {\bibinfo  {journal} {Applied Physics Letters}\ }\textbf {\bibinfo {volume} {107}},\ \bibinfo {pages} {162602} (\bibinfo {year} {2015})}\BibitemShut {NoStop}%
\bibitem [{\citenamefont {Chesca}\ \emph {et~al.}(2021)\citenamefont {Chesca}, \citenamefont {John},\ and\ \citenamefont {Cantor}}]{chesca2021squid}%
  \BibitemOpen
  \bibfield  {author} {\bibinfo {author} {\bibfnamefont {B.}~\bibnamefont {Chesca}}, \bibinfo {author} {\bibfnamefont {D.}~\bibnamefont {John}},\ and\ \bibinfo {author} {\bibfnamefont {R.}~\bibnamefont {Cantor}},\ }\bibfield  {title} {\bibinfo {title} {{SQUID}-arrays coupled to on-chip integrated thin-film superconducting input coils operating coherently},\ }\href@noop {} {\bibfield  {journal} {\bibinfo  {journal} {Applied Physics Letters}\ }\textbf {\bibinfo {volume} {118}},\ \bibinfo {pages} {042601} (\bibinfo {year} {2021})}\BibitemShut {NoStop}%
\bibitem [{\citenamefont {Roy}\ \emph {et~al.}(2015)\citenamefont {Roy}, \citenamefont {Kundu}, \citenamefont {Chand}, \citenamefont {Vadiraj}, \citenamefont {Ranadive}, \citenamefont {Nehra}, \citenamefont {Patankar}, \citenamefont {Aumentado}, \citenamefont {Clerk},\ and\ \citenamefont {Vijay}}]{roy2015broadband}%
  \BibitemOpen
  \bibfield  {author} {\bibinfo {author} {\bibfnamefont {T.}~\bibnamefont {Roy}}, \bibinfo {author} {\bibfnamefont {S.}~\bibnamefont {Kundu}}, \bibinfo {author} {\bibfnamefont {M.}~\bibnamefont {Chand}}, \bibinfo {author} {\bibfnamefont {A.}~\bibnamefont {Vadiraj}}, \bibinfo {author} {\bibfnamefont {A.}~\bibnamefont {Ranadive}}, \bibinfo {author} {\bibfnamefont {N.}~\bibnamefont {Nehra}}, \bibinfo {author} {\bibfnamefont {M.~P.}\ \bibnamefont {Patankar}}, \bibinfo {author} {\bibfnamefont {J.}~\bibnamefont {Aumentado}}, \bibinfo {author} {\bibfnamefont {A.}~\bibnamefont {Clerk}},\ and\ \bibinfo {author} {\bibfnamefont {R.}~\bibnamefont {Vijay}},\ }\bibfield  {title} {\bibinfo {title} {Broadband parametric amplification with impedance engineering: Beyond the gain-bandwidth product},\ }\href@noop {} {\bibfield  {journal} {\bibinfo  {journal} {Applied Physics Letters}\ }\textbf {\bibinfo {volume} {107}},\ \bibinfo {pages} {262601} (\bibinfo {year} {2015})}\BibitemShut {NoStop}%
\bibitem [{\citenamefont {Westig}\ and\ \citenamefont {Klapwijk}(2018)}]{westig2018josephson}%
  \BibitemOpen
  \bibfield  {author} {\bibinfo {author} {\bibfnamefont {M.~P.}\ \bibnamefont {Westig}}\ and\ \bibinfo {author} {\bibfnamefont {T.~M.}\ \bibnamefont {Klapwijk}},\ }\bibfield  {title} {\bibinfo {title} {{Josephson} parametric reflection amplifier with integrated directionality},\ }\href@noop {} {\bibfield  {journal} {\bibinfo  {journal} {Physical Review Applied}\ }\textbf {\bibinfo {volume} {9}},\ \bibinfo {pages} {064010} (\bibinfo {year} {2018})}\BibitemShut {NoStop}%
\bibitem [{\citenamefont {Getsinger}(1963)}]{getsinger1963prototypes}%
  \BibitemOpen
  \bibfield  {author} {\bibinfo {author} {\bibfnamefont {W.~J.}\ \bibnamefont {Getsinger}},\ }\bibfield  {title} {\bibinfo {title} {Prototypes for use in broadbanding reflection amplifiers},\ }\href@noop {} {\bibfield  {journal} {\bibinfo  {journal} {IEEE Transactions on Microwave Theory and Techniques}\ }\textbf {\bibinfo {volume} {11}},\ \bibinfo {pages} {486} (\bibinfo {year} {1963})}\BibitemShut {NoStop}%
\bibitem [{\citenamefont {Chesca}\ \emph {et~al.}(2004)\citenamefont {Chesca}, \citenamefont {Koelle},\ and\ \citenamefont {Kleiner}}]{chesca2004squid}%
  \BibitemOpen
  \bibfield  {author} {\bibinfo {author} {\bibfnamefont {B.}~\bibnamefont {Chesca}}, \bibinfo {author} {\bibfnamefont {D.}~\bibnamefont {Koelle}},\ and\ \bibinfo {author} {\bibfnamefont {R.}~\bibnamefont {Kleiner}},\ }\bibfield  {title} {\bibinfo {title} {{SQUID} theory},\ }in\ \href@noop {} {\emph {\bibinfo {booktitle} {The {SQUID} Handbook: Fundamentals and Technology of {SQUID}s and {SQUID} Systems}}},\ Vol.~\bibinfo {volume} {1},\ \bibinfo {editor} {edited by\ \bibinfo {editor} {\bibfnamefont {J.}~\bibnamefont {Clarke}}\ and\ \bibinfo {editor} {\bibfnamefont {A.~I.}\ \bibnamefont {Braginski}}}\ (\bibinfo  {publisher} {Wiley-VCH},\ \bibinfo {year} {2004})\ pp.\ \bibinfo {pages} {29--92}\BibitemShut {NoStop}%
\bibitem [{\citenamefont {Krantz}\ \emph {et~al.}(2013)\citenamefont {Krantz}, \citenamefont {Reshitnyk}, \citenamefont {Wustmann}, \citenamefont {Bylander}, \citenamefont {Gustavsson}, \citenamefont {Oliver}, \citenamefont {Duty}, \citenamefont {Shumeiko},\ and\ \citenamefont {Delsing}}]{krantz2013investigation}%
  \BibitemOpen
  \bibfield  {author} {\bibinfo {author} {\bibfnamefont {P.}~\bibnamefont {Krantz}}, \bibinfo {author} {\bibfnamefont {Y.}~\bibnamefont {Reshitnyk}}, \bibinfo {author} {\bibfnamefont {W.}~\bibnamefont {Wustmann}}, \bibinfo {author} {\bibfnamefont {J.}~\bibnamefont {Bylander}}, \bibinfo {author} {\bibfnamefont {S.}~\bibnamefont {Gustavsson}}, \bibinfo {author} {\bibfnamefont {W.~D.}\ \bibnamefont {Oliver}}, \bibinfo {author} {\bibfnamefont {T.}~\bibnamefont {Duty}}, \bibinfo {author} {\bibfnamefont {V.}~\bibnamefont {Shumeiko}},\ and\ \bibinfo {author} {\bibfnamefont {P.}~\bibnamefont {Delsing}},\ }\bibfield  {title} {\bibinfo {title} {Investigation of nonlinear effects in {Josephson} parametric oscillators used in circuit quantum electrodynamics},\ }\href@noop {} {\bibfield  {journal} {\bibinfo  {journal} {New Journal of Physics}\ }\textbf {\bibinfo {volume} {15}},\ \bibinfo {pages} {105002} (\bibinfo {year} {2013})}\BibitemShut {NoStop}%
\end{thebibliography}%
\end{document}